\documentclass[%
reprint,
superscriptaddress,
amsmath,amssymb,
prb,
]{revtex4-1}

\usepackage{graphicx}
\usepackage{dcolumn}
\usepackage{bm}
\usepackage{multirow}
\usepackage{amsmath,amssymb}
\usepackage{soul}
\usepackage{hyperref}
\usepackage{xcolor}
\hypersetup{
   colorlinks=true,
   linkcolor=blue,
   citecolor=blue,
   urlcolor=blue,
}

\begin{document}

\preprint{APS/123-QED}

\title{Direct observation of magnetic droplet solitons in all-perpendicular spin torque nano-oscillators}

\author{Sunjae~Chung}
\thanks{These two authors contributed equally}
\affiliation{Department of Physics, University of Gothenburg, 412 96 Gothenburg, Sweden}
\affiliation{Department of Applied Physics, School of Engineering Sciences, KTH Royal Institute of Technology, 164 40 Kista, Sweden}
\affiliation{Department of Physics and Astronomy, University Uppsala, 751 20 Uppsala, Sweden}

\author{Q.~Tuan~Le}
\thanks{These two authors contributed equally}
\affiliation{Department of Physics, University of Gothenburg, 412 96 Gothenburg, Sweden}
\affiliation{Department of Applied Physics, School of Engineering Sciences, KTH Royal Institute of Technology, 164 40 Kista, Sweden}

\author{Martina~Ahlberg}
\affiliation{Department of Physics, University of Gothenburg, 412 96 Gothenburg, Sweden}
\affiliation{NanOsc AB, 164 40 Kista, Sweden}

\author{Markus Weigand}
\affiliation{Max Planck Institute for Intelligent Systems, Stuttgart, Germany}

\author{Iuliia Bykova}
\affiliation{Max Planck Institute for Intelligent Systems, Stuttgart, Germany}

\author{Ahmad A. Awad}
\affiliation{Department of Physics, University of Gothenburg, 412 96 Gothenburg, Sweden}
\affiliation{NanOsc AB, 164 40 Kista, Sweden}

\author{Hamid Mazraati}
\affiliation{Department of Applied Physics, School of Engineering Sciences, KTH Royal Institute of Technology, 164 40 Kista, Sweden}
\affiliation{NanOsc AB, 164 40 Kista, Sweden}

\author{Afshin~Houshang}
\affiliation{Department of Physics, University of Gothenburg, 412 96 Gothenburg, Sweden}
\affiliation{NanOsc AB, 164 40 Kista, Sweden}

\author{Sheng~Jiang}
\affiliation{Department of Applied Physics, School of Engineering Sciences, KTH Royal Institute of Technology, 164 40 Kista, Sweden}

\author{T.~N.~Anh~Nguyen}
\affiliation{Department of Physics, University of Gothenburg, 412 96 Gothenburg, Sweden}
\affiliation{Department of Applied Physics, School of Engineering Sciences, KTH Royal Institute of Technology, 164 40 Kista, Sweden}
\affiliation{Laboratory of Magnetism and Superconductivity, Institute of Materials Science, Vietnam Academy of Science and Technology, 18 Hoang Quoc Viet, Hanoi, Vietnam}

\author{Eberhard Goering}
\affiliation{Max Planck Institute for Intelligent Systems, Stuttgart, Germany}

\author{Gisela Sch\"{u}tz}
\affiliation{Max Planck Institute for Intelligent Systems, Stuttgart, Germany}

\author{Joachim Gr\"{a}fe}
\affiliation{Max Planck Institute for Intelligent Systems, Stuttgart, Germany}

\author{Johan~\AA{}kerman}
\affiliation{Department of Physics, University of Gothenburg, 412 96 Gothenburg, Sweden}
\affiliation{Department of Applied Physics, School of Engineering Sciences, KTH Royal Institute of Technology, 164 40 Kista, Sweden}
\affiliation{NanOsc AB, 164 40 Kista, Sweden}

\begin{abstract}
Magnetic droplets are non-topological dynamical solitons that can be nucleated and sustained in nano-contact based spin torque nano-oscillators (NC-STNOs) with perpendicular anisotropy free layers. While originally predicted in
all-perpendicular NC-STNOs, all experimental demonstrations have so far relied on orthogonal devices with an in-plane polarizing layer that requires a strong magnetic field for droplet nucleation. Here, we instead show the nucleation and sustained operation of magnetic droplets in all-perpendicular NC-STNOs in modest perpendicular fields and over a wide range of nano-contact size. The droplet is observed electrically as an intermediate resistance state accompanied by broadband low-frequency microwave noise. Using canted fields, which introduce a non-zero relative angle between the free and fixed layer, the actual droplet precession frequency can also be determined. Finally, the droplet size, its perimeter width, and its fully reversed core are directly observed underneath a 80 nm diameter nano-contact using scanning transmission x-ray microscopy on both the Ni and Co edges. The droplet diameter is 150 nm, \emph{i.e.}~almost twice the nominal size of the nano-contact, and the droplet has a perimeter width of about 70 nm.
\end{abstract}

\flushbottom
\maketitle

\thispagestyle{empty}

\section*{Introduction}
Non-topological magnetodynamical solitons\cite{Braun2012}, such as droplets\cite{Ivanov1977, Hoefer2010, Iacocca2014, Mohseni2013, Macia2014, Mohseni2013b, Chung2014, Lendinez2015, Chung2015, Chung2016, Xiao2016, Lendinez2017prapplied,Slobodianiuk2017jmmm} and spin wave (SW) bullets\cite{Slavin2005,Bonetti2010,Demidov2010,Demidov2012,Bonetti2012,Dumas2013}, are condensed states of SWs deriving their stability from the intrinsic precession of their spins. For their nucleation, these condensed states require a high local SW density, which in metals can only be achieved using highly focused spin currents\cite{Slonczewski1996,Berger1996,Slonczewski1999}.
In contrast to topological static solitons, such as vortices\cite{Raabe2000,Shinjo2000} and skyrmions\cite{Rossler2006,Pfleiderer2009,Yu2010b,Romming2013},
the non-topological dynamical solitons will dissipate from SW damping if the spin current is removed. The dynamics can also stabilize a  topological state that otherwise would not be stable, an example being the dynamical skyrmion\cite{Zhou2015,Liu2015b}.

High spin current densities can be achieved in so-called nano-contact spin torque nano-oscillators (STNOs from hereon) where a charge current is injected into an extended GMR trilayer through a nano-contact\cite{Dumas2014, Chen2016procieee}. Historically, STNOs were fabricated with all the constituent magnetic layers having in-plane remanent states.\cite{tsoi1998prl,tsoi2000nt,rippard2004prl} In this system, the magnetodynamic non-linearity\cite{SlavinKabos2005,Slavin2009} favors propagating SWs\cite{Slonczewski1999,Bonetti2010,Bonetti2012} when the free layer magnetization is saturated towards the film normal (positive non-linearity), and SW bullets\cite{Slavin2005,Bonetti2010,Bonetti2012} when it is saturated towards the plane (negative non-linearity). Propagating SWs, in particular in the form of SW beams\cite{Hoefer2008prb,Madami2015prb,Houshang2016natnano}, are \emph{e.g.} crucial for the synchronization of multiple STNOs\cite{kaka2005nt,mancoff2005nt,sani2013ntc,Houshang2016natnano}. In STNOs where the free layer has a large perpendicular magnetic anisotropy (PMA) the non-linearity is negative at all fields and any auto-oscillation is inherently self-localized underneath the nano-contact, which promotes a high local SW density.\cite{Rippard2010prb,Mohseni2011} At a critical SW density, the magnetodynamics can then condense into a magnetic droplet soliton, which is characterized by a largely reversed core and a perimeter where all spins precess with the same frequency, and, in ideal conditions, with the same phase.\cite{Ivanov1977,Hoefer2010}

All experimental realizations of droplets have so far relied on so-called orthogonal spin valves, where the fixed layer magnetization has an easy-plane orientation (\emph{e.g.} Co or NiFe).\cite{Mohseni2013,Macia2014,Chung2014, Lendinez2015, Chung2015, Chung2016, Lendinez2017prapplied} To nucleate a droplet, a perpendicular field has to be applied, which tilts, or saturates, the fixed layer out-of-plane. The combination of a tilted fixed layer magnetization and a large Oersted field from the drive current modifies the effective magnetic field landscape in such a way that the droplet experiences a so-called drift instability\cite{Hoefer2010,Lendinez2015},
\emph{i.e.} it may leave the nano-contact region and dissipate out, after which a new droplet can form. The drift instability complicates the experimental characterization of the intrinsic properties of the droplet. As a recent example, attempts at determining the degree of reversal of the droplet core using scanning transmission x-ray microscopy (STXM) resulted in much smaller estimates ($\approx 25^\circ$) than theoretically predicted ($\approx 180^\circ$), and an apparent non-circular droplet shape.\cite{Backes2015} It would therefore be highly valuable to realize, and directly observe, droplets in less asymmetric STNOs. In this work we realize magnetic droplets in STNOs based on all-perpendicular spin valves and use both electrical and STXM measurements to study their properties. Our electrical measurements indicate a much more stable droplet in perpendicular fields than in tilted fields. Using both the Ni and Co edges, our STXM results show that the droplet core is essentially completely reversed ($\approx 180^\circ$) and that the droplet has a highly circular shape.

\begin{figure}[t]
   \centering
\includegraphics [width=3.4in]{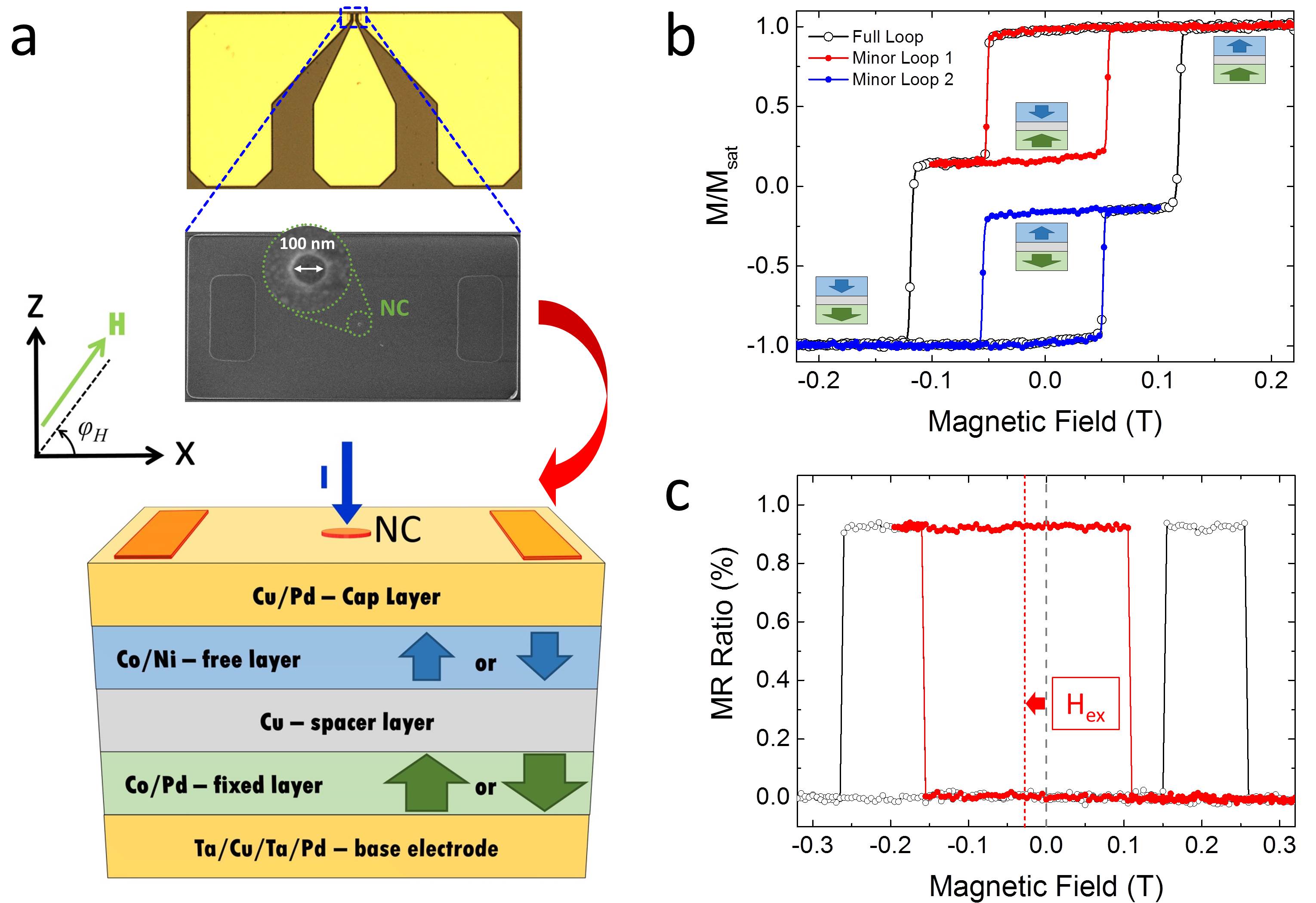}
\centering
\caption{ \label{figure1} \textbf{Device structure. Magnetic and electrical characterization.} (\textbf{a}) Schematic of an all-perpendicular NC-STO composed of Co/Pd (fixed) and Co/Ni (free) multilayers with a Cu spacer. The current $I_{dc}$ flows through the nanocontact (NC) fabricated on top of the stack.
The magnetic field $H$ is applied at an angle $\varphi_{H}$ from the film plane. (\textbf{b}) Full (black circles) and minor (red and blue dots) hysteresis loops of the unpatterned material stack in a perpendicular field, showing entirely decoupled switching of the free and fixed layers before processing. (\textbf{c}) Full (black circles) and minor loop (red dots) low-current ($I_{dc}$~=~-0.6~mA) magnetoresistance (MR) measurements of the patterned NC-STO showing MR of about 1\% and some process induced interlayer coupling of about -0.03~T.}
\end{figure}

\section*{Results}
Fig.1a shows a schematic of the type of all-perpendicular STNO studied in this work, having a Co/Pd multilayer fixed layer and a Co/Ni multilayer free layer, both with sufficient perpendicular magnetic anisotropy (PMA) to have their remanent states along the film normal. The drive current is provided through a nano-contact with diameters ranging from 50 to 150 nm. Fig.1b shows major and minor magnetization hysteresis loops of the full unpatterned material stack with two distinct switching fields corresponding to the fixed and free layer respectively. The symmetry of the minor loops indicate negligible coupling between the fixed and the free layer before patterning. Fig.1c shows a magnetoresistance (MR) hysteresis loop of a fully processed STNO having about 1\% MR and about 0.03 T interlayer coupling after patterning.

Fig.2a shows the resistance variation of a 100 nm nano-contact STNO as the drive current is swept back and forth at three different perpendicular field strengths. At a negative current of about -12 mA and in a field of 0.25 T, there is a sharp step in resistance indicating the nucleation or collapse of a droplet depending on the current sweep direction. The step value is about 60\% of the total difference between the P and the AP states, consistent with a droplet, and its location moves linearly to higher current magnitudes if the field is increased, consistent with the stiffening of the SWs and the field dependence of the Slonczewski threshold current for a spin transfer torque driven SW instability\cite{Slonczewski1999}. The inset shows the same resistance step after the subtraction of the shared parabolic background at all fields. A further direct indication of a droplet is the appearance of broadband microwave noise at low frequency, only observed in the intermediate resistance state, which arises due to the particle-like Brownian motion of the droplet underneath the nano-contact\cite{Hoefer2010,Mohseni2013,Chung2016}.

\begin{figure}
\includegraphics [width=3.4in]{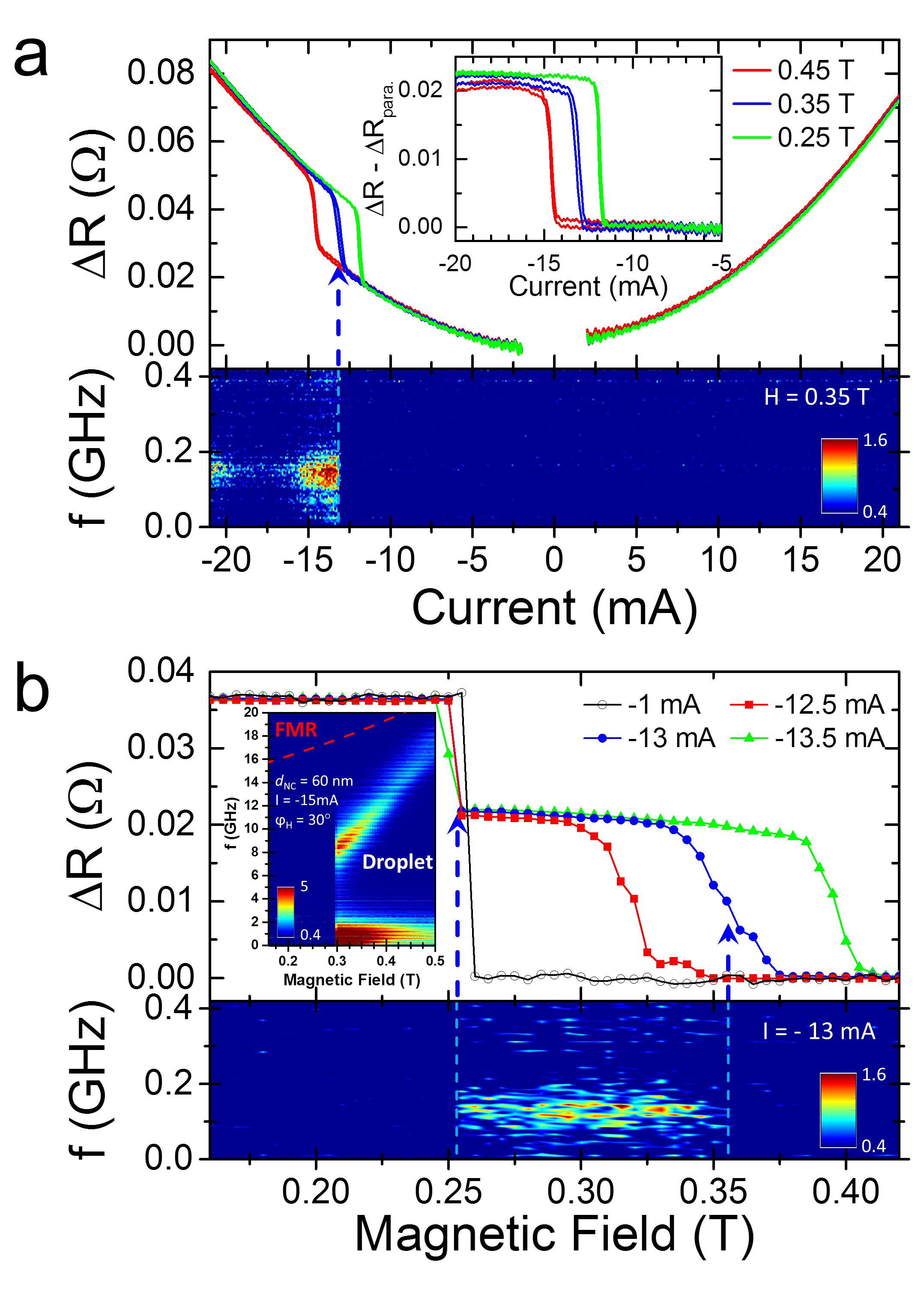}
\centering
\caption{ \label{figure2} \textbf{Droplet nucleation and precession. } (\textbf{a}) Change in the resistance of a 100 nm nano-contact vs.~drive current showing a background of Joule heating and the nucleation of a droplet at a field dependent negative current. The inset shows the same data after subtraction of the parabolic background. Below the resistance measurement is a plot of the power spectral density measured up to 0.4 GHz in a field of 0.35 T, showing how the resistance step is accompanied by the appearance low-frequency microwave noise. (\textbf{b}) Field-sweep resistance measurements from 0.1 to 0.42 T at four different negative currents for the same nano-contact as in (\textbf{a}). At a small negative current (-1 mA) the state switches directly from AP to P at about 0.25 T. For the three large negative currents, the AP state first switches to an intermediate resistance state consistent with a droplet, before gradually switching to the P state. The formation of the droplet is again accompanied by substantial microwave noise. The inset shows a power spectral density measurement taken in a field tilted 30 degrees from the field plane clearly showing both the precession frequency and the microwave noise of the droplet.}
\end{figure}

Fig.2b shows the droplet nucleation as the field is increased from 0.16 to 0.42 T at three different strong negative currents and compared to a field sweep at much lower current. Again, the droplet is characterized by an intermediate resistance value, which first decreases slowly with field until it drops more rapidly towards the resistance value of the P state. The gradual collapse indicates mode hopping between the droplet and the P state. Just as in Fig.2a, the droplet is again accompanied by the appearance of broadband low-frequency microwave noise.

In this all-perpendicular geometry, where the two magnetizations and also the applied magnetic field are all aligned along the film normal, the high-frequency precession of the droplet does not generate any microwave signal since the projection of the precessing spins onto the fixed layer magnetization remains constant in time. We can however prove that the droplet precesses by tilting the applied field closer to the film plane, since this creates a substantial non-collinearity between the free and the fixed layer magnetizations and hence a time-dependent variation of the STNO resistance. The inset shows a microwave measurement as a field applied at 30 degrees switches the STNO from its AP state to a droplet state at about 0.3 T, resulting in a strong signal at 8 GHz, which increases linearly with field strength; the droplet nucleation is again accompanied by substantial microwave noise between 0 and 2 GHz. The precession frequency is deep into the SW gap of the free layer magnetization, consistent with an essentially fully reversed droplet.

It is noteworthy that the low-frequency noise is dramatically higher in tilted fields (inset) compared to perpendicular fields. If the noise is generated by the drop instability, every droplet leaving the NC generates a similar voltage spike in both cases, and the total microwave noise power is hence a good measure of the droplet stability. We are hence lead to conclude that the droplet is highly stable in the perpendicular case.

\begin{figure}
\includegraphics [width=3.4in]{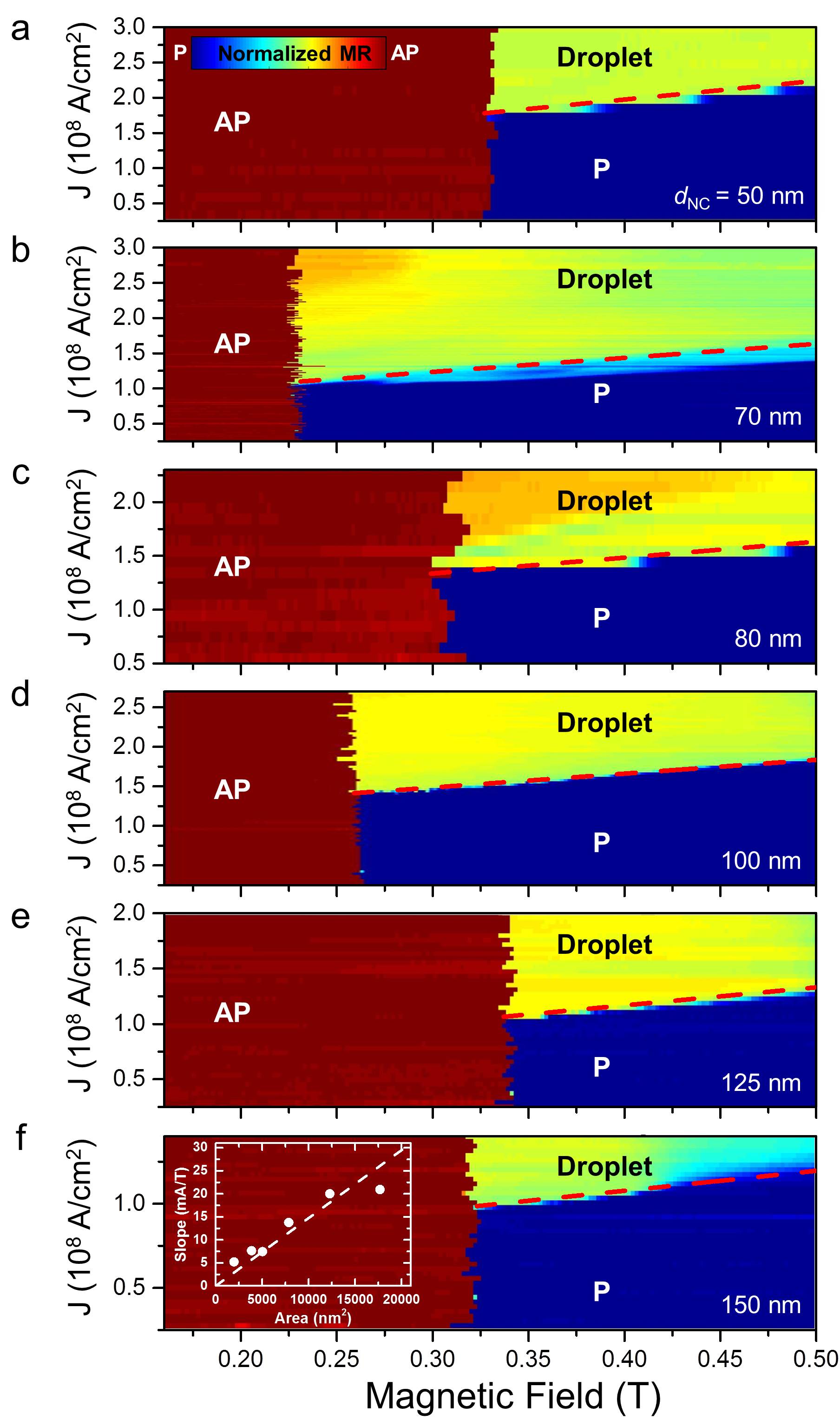}
\centering
\caption{ \label{figure3} \textbf{Droplet nucleation phase diagram.} Field-sweep resistance measurements from 0.16 to 0.5 T at different negative currents for nano-contact diameters ranging from 50 to 150 nm, plotted on a color scale defined by the P and AP states. The droplet state is seen as an intermediate resistance state. The dashed red line marks the linear current-field droplet nucleation boundary. The inset plots the slope of this boundary vs.~nano-contact area together with a linear fit (dashed white line).}
\end{figure}

We have reproduced this general droplet behavior in a large number of STNOs having different nano-contact diameters. Fig.3 shows the corresponding current density/field phase diagram of the free layer magnetization in six different STNOs with diameters ranging from 50 to 150 nm, as measured by the normalized STNO resistance. In all STNOs the AP state either switches into the P state at low current magnitudes, or into the droplet at high current magnitudes, and the droplet nucleation boundary shows a linear dependence on current and field. The ratio of the droplet resistance vs. either the P or the AP resistance, does not depend systematically on the nano-contact diameter; it is rather dominated by device to device variations. The linear slope of the droplet nucleation boundary is similarly independent on nano-contact diameter as the current density determines the nucleation. The inset in the bottom sub figure of Fig.3 shows this slope expressed as current per field and plotted vs. the nano-contact area, again confirming that the current density governs the nucleation.\par

We finally turn to our scanning transmission x-ray microscopy (STXM) measurements on an 80 nm NC. Fig.4a shows a spatial map of the $m_z$ component of the Ni moments normalized to the up and down states well outside of the droplet region. The left inset shows a 3D-rendered cross-section of the same data. The map reveals an essentially fully reversed droplet core with a well defined circular shape. The droplet diameter is approximately 150 nm, \emph{i.e.}~almost twice as large as the nominal NC diameter, and the droplet perimeter width (10-90\%) is about 70 nm. Fig.4b shows the same analysis using the Co moment. As we have signal from Co moments both in the free and the fixed layer and the normalization procedure switches both the free and the fixed layers, a second normalization step was done using the relative Co content in the free and the fixed layers respectively.\par
The spatial map of the $m_z$ component of the free layer Co moments corroborates the conclusions drawn from the Ni signal, such as a fully reversed core, as well as the diameter and perimeter width values. In addition, we find that the minor deviations of the perimeter from being a perfect circle are  uncorrelated between the Ni and Co maps. These deviations can therefore be ascribed to measurement noise and are not intrinsic to the droplet. The droplet is hence even more circular than what the individual maps would indicate on their own.  We can hence fit circles to the $(x,y)$ position of all data with a certain $m_z$ value and this way trace out the droplet perimeter with greater accuracy. The resulting droplet envelope is shown in the right inset in Fig.4a.
\begin{figure}
\includegraphics [width=3.4in]{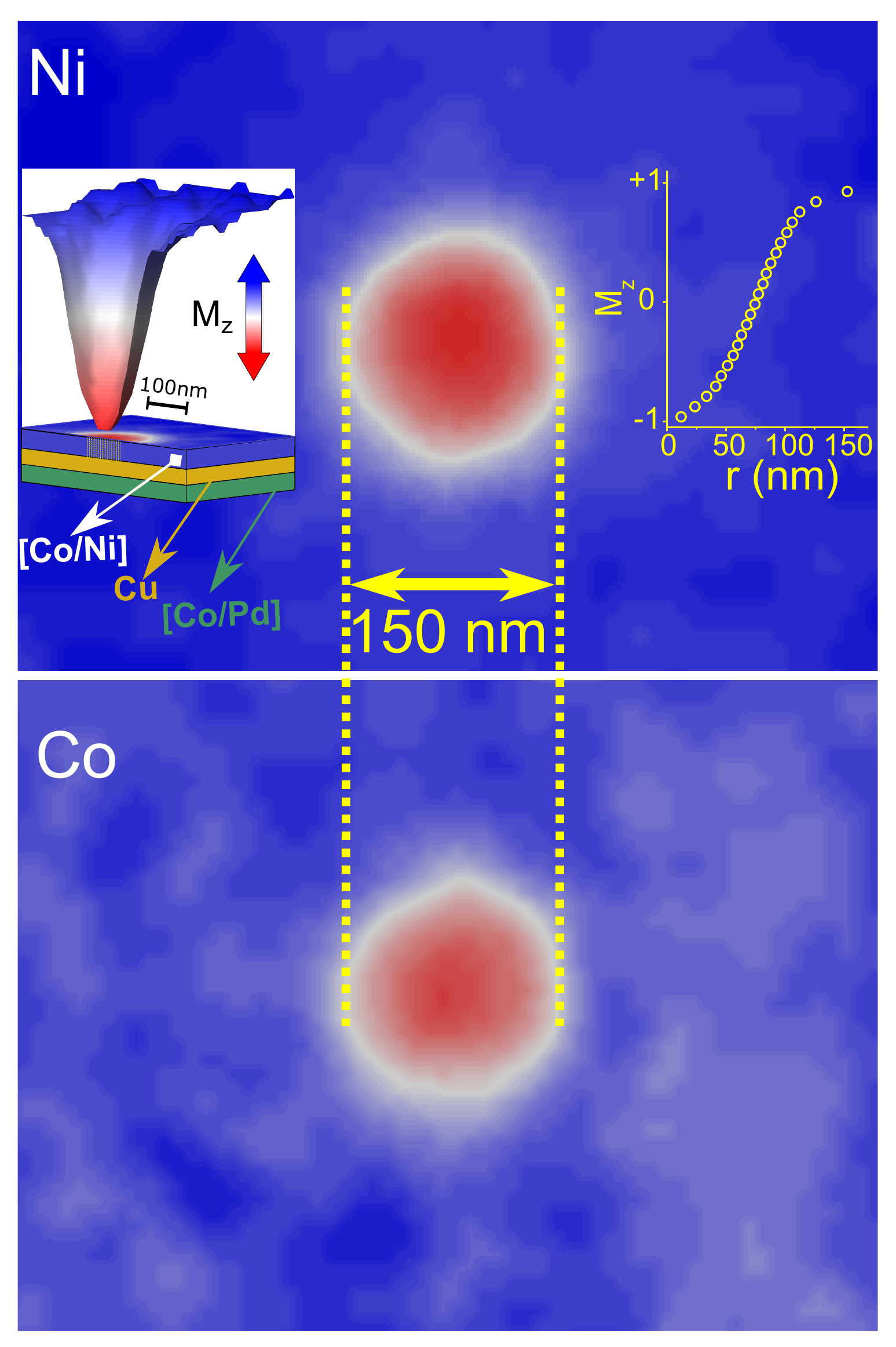}
\centering
\caption{ \label{figure4} \textbf{Direct STXM observation of a reversed droplet.}
Spatial map of the $m_z$ component of the Ni (top) and Co (bottom) moments of the free layer. Both the Ni and Co data reveal a fully reversed droplet with a diameter of about 150 nm. The left inset shows a 3D rendered cross section of the Ni STXM data. The right inset shows the detailed droplet perimeter profile extracted from the Ni data assuming a circularly symmetric droplet.}
\end{figure}
It is noteworthy that our direct measurement of the droplet diameter yields a much larger droplet than predicted by theory and micromagnetic simulations\cite{Hoefer2010}. However, these simulations assume a perfect cylindrical current distribution underneath the NC, whereas recent experimental and numerical result indicate a large lateral current spread underneath the NC\cite{Banuazizi2017nanoscale} resulting in both spin transfer torque over an area greater than the NC and substantial in-plane spin transfer torque acting on the droplet perimeter, which could potentially also affect its size. While it is well beyond the scope of our study to further elucidate these effects, they highlight the need for further modelling of how a realistic three-dimensional current distribution underneath the NC affects the droplet size, perimeter width, and even stability.
\section*{Conclusion}
As any significant drift instability\cite{Hoefer2010,Lendinez2015} of the droplet would likely have perturbed its apparent shape\cite{Backes2015} our results indicate that droplets in all-perpendicular STNOs are both fully reversed and highly stable. The realization of stable room-temperature droplets not suffering from drift instability is crucial for their further studies. In addition, the all-perpendicular geometry, in contrast to the orthogonal, can be easily realized using magnetic tunnel junctions (MTJs) with and without substantial Dzyaloshinskii–Moriya interaction. Our demonstration of stable droplets in all-perpendicular spin valve STNOs hence represents the first step towards utilizing droplets, and potentially dynamical skyrmions\cite{Zhou2015}, in high-output MTJ based STNOs.
\section*{methods}
\textbf{Sample Preparation} A full stack composed of a Ta~($4$~nm)/ Cu~($14$~nm) / Ta~($4$~nm) / Pd~($2$~nm) seed layer and an all-perpendicular pseudo-spin valve [Co~($0.35$~nm) / Pd~($0.7$~nm)]${\times5}$ / Co~($0.35$~nm) / Cu~($5$~nm) / [Co~($0.22$~nm) / Ni~($0.68$~nm)]${\times4}$ / Co~($0.22$~nm), capped by a Cu~($2$~nm) / Pd~($2$~nm) layer, was deposited on a thermally oxidized Si wafer by magnetron sputtering technique (numbers in parentheses are thicknesses in nanometers). Using a combination of optical lithography and etching techniques, 8 $\mu$m $\times$ 16 $\mu$m mesas were fabricated on the stack wafer and insulated by a 30-nm-thick SiO$_2$ film using chemical vapor deposition (CVD). Electron beam lithography was used to pattern nanocontacts, with circular sizes varying from 50 to 150 nm in diameter, on top of each mesa. SiO$_2$ was then etched through by the reactive ion etching (RIE) technique to open the contacts. The NC-STO device fabrication was completed by the deposition of Cu $500$~nm / Au $100$~nm top electrode and lift-off processing. For STXM measurements, the similar stack deposition and processing were employed to fabricate NC-STOs on 300-nm-thick LPCVD silicon nitride Si wafer, then the highly selective deep RIE was used to remove Si from backside of the device wafer and leave nitride membranes underneath NC-STOs for X-ray illumination.\par

\textbf{Magnetic and Electrical Characterization} Alternating Gradient Magnetometry was used to measure the magnetization hysteresis loops of the unpatterned material stacks. \textit{dc} and microwave measurements of the fabricated STOs were carried out using our custom-built setup, which allows the manipulation of magnetic field strength, polarity, and angle. The electromagnet can generate a field between -0.5 to +0.5 T and its rotational base easily controls the field angle  between 0 and 90$^{\circ}$. The device is connected by a ground--signal--ground probe to a \textit{dc}-current source (Keithley 6221), a nanovoltmeter (Keithley 2182A), and a spectrum analyzer (R \& S FSQ26). A 0--40GHz bias-tee is used to separate the bias input and the generated microwave signal. The latter is amplified by a low-noise amplifier (operational range: 0.1--26.5 GHz) before being sent to the spectrum analyzer.\par

\textbf{Scanning transmission x-ray microscopy} STXM measurements were conducted at the MPI~IS operated MAXYMUS end station at the UE46-PGM2 beam line at the BESSY II synchrotron radiation facility. The samples were illuminated under normal incidence by circularly polarized light in an applied out-of-plane field of up to 240 mT that was generated by a set of four rotatable permanent magnets\cite{Nolle2012}. The photon energy was set to the absorption maximum of the Ni $L_3$ and Co $L_3$ edge to get optimal XMCD contrast for imaging of each element. Intensities were locally averaged over the nominal resolution of the focusing zone plate of 18 nm using a Gaussian filter in ImageJ\cite{ImageJ}.
Magnetization angles from XMCD measurements were calibrated to the saturation magnetization of the free layer. Both external field and photon polarization were flipped to compensate for intensity variations of the x-ray beam. The noise level in the reference measurements determines the uncertainty of the subsequent XMCD measurements; a spin angle larger than 160$^\circ$, \emph{i.e.}~within 20$^\circ$ of full reversal, is considered fully reversed. A lock-in-like data acquisition scheme based on an avalanche photo diode and a custom FPGA system allows ultra low-noise measurements.

This work was partially supported by the ERC Starting Grant 307144 ``Mustang'', the Swedish Foundation for Strategic Research (SSF) Successful Research Leaders program, the Swedish Research Council (VR), the Göran Gustafsson foundation, and the Knut and Alice Wallenberg Foundation. Helmholtz Zentrum Berlin is acknowledged for allocating beam time at the BESSY II synchrotron radiation facility. Financial support by the Baden-Württemberg Stiftung in the framework of the Kompetenznetz Funktionelle Nanostrukturen is gratefully acknowledged.

S.C. and T.Q.L. performed the electrical measurements. S.C., T.Q.L., S.J., A.H. and T.N.A.N. fabricated the devices.
M.A., J.G, M.W. and I.B carried out the STXM measurements. J.\AA{}. coordinated the project. All authors analyzed the results and co-wrote the manuscript. Correspondence and requests for materials should be addressed to J. \AA{}kerman (email: johan.akerman@physics.gu.se).

\bibliography{AllPerpDroplet_APS}

\begin{thebibliography}{50}%
\makeatletter
\providecommand \@ifxundefined [1]{%
 \@ifx{#1\undefined}
}%
\providecommand \@ifnum [1]{%
 \ifnum #1\expandafter \@firstoftwo
 \else \expandafter \@secondoftwo
 \fi
}%
\providecommand \@ifx [1]{%
 \ifx #1\expandafter \@firstoftwo
 \else \expandafter \@secondoftwo
 \fi
}%
\providecommand \natexlab [1]{#1}%
\providecommand \enquote  [1]{``#1''}%
\providecommand \bibnamefont  [1]{#1}%
\providecommand \bibfnamefont [1]{#1}%
\providecommand \citenamefont [1]{#1}%
\providecommand \href@noop [0]{\@secondoftwo}%
\providecommand \href [0]{\begingroup \@sanitize@url \@href}%
\providecommand \@href[1]{\@@startlink{#1}\@@href}%
\providecommand \@@href[1]{\endgroup#1\@@endlink}%
\providecommand \@sanitize@url [0]{\catcode `\\12\catcode `\$12\catcode
  `\&12\catcode `\#12\catcode `\^12\catcode `\_12\catcode `\%12\relax}%
\providecommand \@@startlink[1]{}%
\providecommand \@@endlink[0]{}%
\providecommand \url  [0]{\begingroup\@sanitize@url \@url }%
\providecommand \@url [1]{\endgroup\@href {#1}{\urlprefix }}%
\providecommand \urlprefix  [0]{URL }%
\providecommand \Eprint [0]{\href }%
\providecommand \doibase [0]{http://dx.doi.org/}%
\providecommand \selectlanguage [0]{\@gobble}%
\providecommand \bibinfo  [0]{\@secondoftwo}%
\providecommand \bibfield  [0]{\@secondoftwo}%
\providecommand \translation [1]{[#1]}%
\providecommand \BibitemOpen [0]{}%
\providecommand \bibitemStop [0]{}%
\providecommand \bibitemNoStop [0]{.\EOS\space}%
\providecommand \EOS [0]{\spacefactor3000\relax}%
\providecommand \BibitemShut  [1]{\csname bibitem#1\endcsname}%
\let\auto@bib@innerbib\@empty
\bibitem [{\citenamefont {Braun}(2012)}]{Braun2012}%
  \BibitemOpen
  \bibfield  {author} {\bibinfo {author} {\bibfnamefont {H.-B.}\ \bibnamefont
  {Braun}},\ }\href {\doibase 10.1080/00018732.2012.663070} {\bibfield
  {journal} {\bibinfo  {journal} {Adv. Phys.}\ }\textbf {\bibinfo {volume}
  {61}},\ \bibinfo {pages} {1} (\bibinfo {year} {2012})}\BibitemShut {NoStop}%
\bibitem [{\citenamefont {Ivanov}\ and\ \citenamefont
  {Kosevich}(1977)}]{Ivanov1977}%
  \BibitemOpen
  \bibfield  {author} {\bibinfo {author} {\bibfnamefont {B.}~\bibnamefont
  {Ivanov}}\ and\ \bibinfo {author} {\bibfnamefont {A.}~\bibnamefont
  {Kosevich}},\ }\href@noop {} {\bibfield  {journal} {\bibinfo  {journal} {Zh.
  Eksp. Teor. Fiz.}\ }\textbf {\bibinfo {volume} {72}},\ \bibinfo {pages}
  {2000} (\bibinfo {year} {1977})}\BibitemShut {NoStop}%
\bibitem [{\citenamefont {Hoefer}\ \emph {et~al.}(2010)\citenamefont {Hoefer},
  \citenamefont {Silva},\ and\ \citenamefont {Keller}}]{Hoefer2010}%
  \BibitemOpen
  \bibfield  {author} {\bibinfo {author} {\bibfnamefont {M.~A.}\ \bibnamefont
  {Hoefer}}, \bibinfo {author} {\bibfnamefont {T.~J.}\ \bibnamefont {Silva}}, \
  and\ \bibinfo {author} {\bibfnamefont {M.~W.}\ \bibnamefont {Keller}},\
  }\href {\doibase 10.1103/PhysRevB.82.054432} {\bibfield  {journal} {\bibinfo
  {journal} {Phys. Rev. B}\ }\textbf {\bibinfo {volume} {82}},\ \bibinfo
  {pages} {054432} (\bibinfo {year} {2010})}\BibitemShut {NoStop}%
\bibitem [{\citenamefont {Iacocca}\ \emph {et~al.}(2014)\citenamefont
  {Iacocca}, \citenamefont {Dumas}, \citenamefont {Bookman}, \citenamefont
  {Mohseni}, \citenamefont {Chung}, \citenamefont {Hoefer},\ and\ \citenamefont
  {\AA{}kerman}}]{Iacocca2014}%
  \BibitemOpen
  \bibfield  {author} {\bibinfo {author} {\bibfnamefont {E.}~\bibnamefont
  {Iacocca}}, \bibinfo {author} {\bibfnamefont {R.~K.}\ \bibnamefont {Dumas}},
  \bibinfo {author} {\bibfnamefont {L.}~\bibnamefont {Bookman}}, \bibinfo
  {author} {\bibfnamefont {M.}~\bibnamefont {Mohseni}}, \bibinfo {author}
  {\bibfnamefont {S.}~\bibnamefont {Chung}}, \bibinfo {author} {\bibfnamefont
  {M.~A.}\ \bibnamefont {Hoefer}}, \ and\ \bibinfo {author} {\bibfnamefont
  {J.}~\bibnamefont {\AA{}kerman}},\ }\href {\doibase
  10.1103/PhysRevLett.112.047201} {\bibfield  {journal} {\bibinfo  {journal}
  {Phys. Rev. Lett.}\ }\textbf {\bibinfo {volume} {112}},\ \bibinfo {pages}
  {047201} (\bibinfo {year} {2014})}\BibitemShut {NoStop}%
\bibitem [{\citenamefont {Mohseni}\ \emph {et~al.}(2013)\citenamefont
  {Mohseni}, \citenamefont {Sani}, \citenamefont {Persson}, \citenamefont
  {Nguyen}, \citenamefont {Chung}, \citenamefont {Pogoryelov}, \citenamefont
  {Muduli}, \citenamefont {Iacocca}, \citenamefont {Eklund}, \citenamefont
  {Dumas}, \citenamefont {Bonetti}, \citenamefont {Deac}, \citenamefont
  {Hoefer},\ and\ \citenamefont {Akerman}}]{Mohseni2013}%
  \BibitemOpen
  \bibfield  {author} {\bibinfo {author} {\bibfnamefont {S.~M.}\ \bibnamefont
  {Mohseni}}, \bibinfo {author} {\bibfnamefont {S.~R.}\ \bibnamefont {Sani}},
  \bibinfo {author} {\bibfnamefont {J.}~\bibnamefont {Persson}}, \bibinfo
  {author} {\bibfnamefont {T.~N.~A.}\ \bibnamefont {Nguyen}}, \bibinfo {author}
  {\bibfnamefont {S.}~\bibnamefont {Chung}}, \bibinfo {author} {\bibfnamefont
  {Y.}~\bibnamefont {Pogoryelov}}, \bibinfo {author} {\bibfnamefont {P.~K.}\
  \bibnamefont {Muduli}}, \bibinfo {author} {\bibfnamefont {E.}~\bibnamefont
  {Iacocca}}, \bibinfo {author} {\bibfnamefont {A.}~\bibnamefont {Eklund}},
  \bibinfo {author} {\bibfnamefont {R.~K.}\ \bibnamefont {Dumas}}, \bibinfo
  {author} {\bibfnamefont {S.}~\bibnamefont {Bonetti}}, \bibinfo {author}
  {\bibfnamefont {A.}~\bibnamefont {Deac}}, \bibinfo {author} {\bibfnamefont
  {M.~a.}\ \bibnamefont {Hoefer}}, \ and\ \bibinfo {author} {\bibfnamefont
  {J.}~\bibnamefont {Akerman}},\ }\href {\doibase 10.1126/science.1230155}
  {\bibfield  {journal} {\bibinfo  {journal} {Science}\ }\textbf {\bibinfo
  {volume} {339}},\ \bibinfo {pages} {1295} (\bibinfo {year}
  {2013})}\BibitemShut {NoStop}%
\bibitem [{\citenamefont {Maci{\`{a}}}\ \emph {et~al.}(2014)\citenamefont
  {Maci{\`{a}}}, \citenamefont {Backes},\ and\ \citenamefont
  {Kent}}]{Macia2014}%
  \BibitemOpen
  \bibfield  {author} {\bibinfo {author} {\bibfnamefont {F.}~\bibnamefont
  {Maci{\`{a}}}}, \bibinfo {author} {\bibfnamefont {D.}~\bibnamefont {Backes}},
  \ and\ \bibinfo {author} {\bibfnamefont {A.~D.}\ \bibnamefont {Kent}},\
  }\href {\doibase 10.1038/nnano.2014.255} {\bibfield  {journal} {\bibinfo
  {journal} {Nat. Nanotechnol}\ }\textbf {\bibinfo {volume} {9}},\ \bibinfo
  {pages} {992} (\bibinfo {year} {2014})}\BibitemShut {NoStop}%
\bibitem [{\citenamefont {Mohseni}\ \emph {et~al.}(2014)\citenamefont
  {Mohseni}, \citenamefont {Sani}, \citenamefont {Dumas}, \citenamefont
  {Persson}, \citenamefont {Nguyen}, \citenamefont {Chung}, \citenamefont
  {Pogoryelov}, \citenamefont {Muduli}, \citenamefont {Iacocca}, \citenamefont
  {Eklund},\ and\ \citenamefont {\AA{}kerman}}]{Mohseni2013b}%
  \BibitemOpen
  \bibfield  {author} {\bibinfo {author} {\bibfnamefont {S.}~\bibnamefont
  {Mohseni}}, \bibinfo {author} {\bibfnamefont {S.}~\bibnamefont {Sani}},
  \bibinfo {author} {\bibfnamefont {R.}~\bibnamefont {Dumas}}, \bibinfo
  {author} {\bibfnamefont {J.}~\bibnamefont {Persson}}, \bibinfo {author}
  {\bibfnamefont {T.~A.}\ \bibnamefont {Nguyen}}, \bibinfo {author}
  {\bibfnamefont {S.}~\bibnamefont {Chung}}, \bibinfo {author} {\bibfnamefont
  {Y.}~\bibnamefont {Pogoryelov}}, \bibinfo {author} {\bibfnamefont
  {P.}~\bibnamefont {Muduli}}, \bibinfo {author} {\bibfnamefont
  {E.}~\bibnamefont {Iacocca}}, \bibinfo {author} {\bibfnamefont
  {A.}~\bibnamefont {Eklund}}, \ and\ \bibinfo {author} {\bibfnamefont
  {J.}~\bibnamefont {\AA{}kerman}},\ }\href {\doibase
  http://dx.doi.org/10.1016/j.physb.2013.10.023} {\bibfield  {journal}
  {\bibinfo  {journal} {Physica B}\ }\textbf {\bibinfo {volume} {435}},\
  \bibinfo {pages} {84 } (\bibinfo {year} {2014})}\BibitemShut {NoStop}%
\bibitem [{\citenamefont {Chung}\ \emph {et~al.}(2014)\citenamefont {Chung},
  \citenamefont {Mohseni}, \citenamefont {Sani}, \citenamefont {Iacocca},
  \citenamefont {Dumas}, \citenamefont {Anh~Nguyen}, \citenamefont
  {Pogoryelov}, \citenamefont {Muduli}, \citenamefont {Eklund}, \citenamefont
  {Hoefer},\ and\ \citenamefont {Åkerman}}]{Chung2014}%
  \BibitemOpen
  \bibfield  {author} {\bibinfo {author} {\bibfnamefont {S.}~\bibnamefont
  {Chung}}, \bibinfo {author} {\bibfnamefont {S.~M.}\ \bibnamefont {Mohseni}},
  \bibinfo {author} {\bibfnamefont {S.~R.}\ \bibnamefont {Sani}}, \bibinfo
  {author} {\bibfnamefont {E.}~\bibnamefont {Iacocca}}, \bibinfo {author}
  {\bibfnamefont {R.~K.}\ \bibnamefont {Dumas}}, \bibinfo {author}
  {\bibfnamefont {T.~N.}\ \bibnamefont {Anh~Nguyen}}, \bibinfo {author}
  {\bibfnamefont {Y.}~\bibnamefont {Pogoryelov}}, \bibinfo {author}
  {\bibfnamefont {P.~K.}\ \bibnamefont {Muduli}}, \bibinfo {author}
  {\bibfnamefont {A.}~\bibnamefont {Eklund}}, \bibinfo {author} {\bibfnamefont
  {M.}~\bibnamefont {Hoefer}}, \ and\ \bibinfo {author} {\bibfnamefont
  {J.}~\bibnamefont {Åkerman}},\ }\href {\doibase
  http://dx.doi.org/10.1063/1.4870696} {\bibfield  {journal} {\bibinfo
  {journal} {J. Appl. Phys.}\ }\textbf {\bibinfo {volume} {115}},\ \bibinfo
  {pages} {172612} (\bibinfo {year} {2014})}\BibitemShut {NoStop}%
\bibitem [{\citenamefont {Lend{\'{i}}nez}\ \emph {et~al.}(2015)\citenamefont
  {Lend{\'{i}}nez}, \citenamefont {Statuto}, \citenamefont {Backes},
  \citenamefont {Kent},\ and\ \citenamefont {Maci{\`{a}}}}]{Lendinez2015}%
  \BibitemOpen
  \bibfield  {author} {\bibinfo {author} {\bibfnamefont {S.}~\bibnamefont
  {Lend{\'{i}}nez}}, \bibinfo {author} {\bibfnamefont {N.}~\bibnamefont
  {Statuto}}, \bibinfo {author} {\bibfnamefont {D.}~\bibnamefont {Backes}},
  \bibinfo {author} {\bibfnamefont {A.~D.}\ \bibnamefont {Kent}}, \ and\
  \bibinfo {author} {\bibfnamefont {F.}~\bibnamefont {Maci{\`{a}}}},\ }\href
  {\doibase 10.1103/PhysRevB.92.174426} {\bibfield  {journal} {\bibinfo
  {journal} {Phys. Rev. B}\ }\textbf {\bibinfo {volume} {92}},\ \bibinfo
  {pages} {174426} (\bibinfo {year} {2015})}\BibitemShut {NoStop}%
\bibitem [{\citenamefont {Chung}\ \emph {et~al.}(2015)\citenamefont {Chung},
  \citenamefont {{Majid Mohseni}}, \citenamefont {Eklund}, \citenamefont
  {D{\"{u}}rrenfeld}, \citenamefont {Ranjbar}, \citenamefont {Sani},
  \citenamefont {{Anh Nguyen}}, \citenamefont {Dumas},\ and\ \citenamefont
  {{\AA}kerman}}]{Chung2015}%
  \BibitemOpen
  \bibfield  {author} {\bibinfo {author} {\bibfnamefont {S.}~\bibnamefont
  {Chung}}, \bibinfo {author} {\bibfnamefont {S.}~\bibnamefont {{Majid
  Mohseni}}}, \bibinfo {author} {\bibfnamefont {A.}~\bibnamefont {Eklund}},
  \bibinfo {author} {\bibfnamefont {P.}~\bibnamefont {D{\"{u}}rrenfeld}},
  \bibinfo {author} {\bibfnamefont {M.}~\bibnamefont {Ranjbar}}, \bibinfo
  {author} {\bibfnamefont {S.~R.}\ \bibnamefont {Sani}}, \bibinfo {author}
  {\bibfnamefont {T.~N.}\ \bibnamefont {{Anh Nguyen}}}, \bibinfo {author}
  {\bibfnamefont {R.~K.}\ \bibnamefont {Dumas}}, \ and\ \bibinfo {author}
  {\bibfnamefont {J.}~\bibnamefont {{\AA}kerman}},\ }\href {\doibase
  10.1063/1.4932358} {\bibfield  {journal} {\bibinfo  {journal} {Low Temp.
  Phys.}\ }\textbf {\bibinfo {volume} {41}},\ \bibinfo {pages} {833} (\bibinfo
  {year} {2015})}\BibitemShut {NoStop}%
\bibitem [{\citenamefont {Chung}\ \emph {et~al.}(2016)\citenamefont {Chung},
  \citenamefont {Eklund}, \citenamefont {Iacocca}, \citenamefont {Mohseni},
  \citenamefont {Sani}, \citenamefont {Bookman}, \citenamefont {Hoefer},
  \citenamefont {Dumas},\ and\ \citenamefont {{\AA}kerman}}]{Chung2016}%
  \BibitemOpen
  \bibfield  {author} {\bibinfo {author} {\bibfnamefont {S.}~\bibnamefont
  {Chung}}, \bibinfo {author} {\bibfnamefont {A.}~\bibnamefont {Eklund}},
  \bibinfo {author} {\bibfnamefont {E.}~\bibnamefont {Iacocca}}, \bibinfo
  {author} {\bibfnamefont {S.~M.}\ \bibnamefont {Mohseni}}, \bibinfo {author}
  {\bibfnamefont {S.~R.}\ \bibnamefont {Sani}}, \bibinfo {author}
  {\bibfnamefont {L.}~\bibnamefont {Bookman}}, \bibinfo {author} {\bibfnamefont
  {M.~A.}\ \bibnamefont {Hoefer}}, \bibinfo {author} {\bibfnamefont {R.~K.}\
  \bibnamefont {Dumas}}, \ and\ \bibinfo {author} {\bibfnamefont
  {J.}~\bibnamefont {{\AA}kerman}},\ }\href {\doibase 10.1038/ncomms11209}
  {\bibfield  {journal} {\bibinfo  {journal} {Nat. Commun.}\ }\textbf {\bibinfo
  {volume} {7}},\ \bibinfo {pages} {11209} (\bibinfo {year}
  {2016})}\BibitemShut {NoStop}%
\bibitem [{\citenamefont {Xiao}\ \emph {et~al.}(2017)\citenamefont {Xiao},
  \citenamefont {Tiberkevich}, \citenamefont {Liu}, \citenamefont {Liu},
  \citenamefont {Mohseni}, \citenamefont {Chung}, \citenamefont {Ahlberg},
  \citenamefont {Slavin}, \citenamefont {{\AA}kerman},\ and\ \citenamefont
  {Zhou}}]{Xiao2016}%
  \BibitemOpen
  \bibfield  {author} {\bibinfo {author} {\bibfnamefont {D.}~\bibnamefont
  {Xiao}}, \bibinfo {author} {\bibfnamefont {V.}~\bibnamefont {Tiberkevich}},
  \bibinfo {author} {\bibfnamefont {Y.~H.}\ \bibnamefont {Liu}}, \bibinfo
  {author} {\bibfnamefont {Y.~W.}\ \bibnamefont {Liu}}, \bibinfo {author}
  {\bibfnamefont {S.~M.}\ \bibnamefont {Mohseni}}, \bibinfo {author}
  {\bibfnamefont {S.}~\bibnamefont {Chung}}, \bibinfo {author} {\bibfnamefont
  {M.}~\bibnamefont {Ahlberg}}, \bibinfo {author} {\bibfnamefont {A.~N.}\
  \bibnamefont {Slavin}}, \bibinfo {author} {\bibfnamefont {J.}~\bibnamefont
  {{\AA}kerman}}, \ and\ \bibinfo {author} {\bibfnamefont {Y.}~\bibnamefont
  {Zhou}},\ }\href {\doibase 10.1103/PhysRevB.95.024106} {\bibfield  {journal}
  {\bibinfo  {journal} {Phys. Rev. B}\ }\textbf {\bibinfo {volume} {95}},\
  \bibinfo {pages} {024106} (\bibinfo {year} {2017})}\BibitemShut {NoStop}%
\bibitem [{\citenamefont {Lend{\'{i}}nez}\ \emph {et~al.}(2017)\citenamefont
  {Lend{\'{i}}nez}, \citenamefont {Hang}, \citenamefont {V{\'{e}}lez},
  \citenamefont {Hern{\'{a}}ndez}, \citenamefont {Backes}, \citenamefont
  {Kent},\ and\ \citenamefont {Maci{\`{a}}}}]{Lendinez2017prapplied}%
  \BibitemOpen
  \bibfield  {author} {\bibinfo {author} {\bibfnamefont {S.}~\bibnamefont
  {Lend{\'{i}}nez}}, \bibinfo {author} {\bibfnamefont {J.}~\bibnamefont
  {Hang}}, \bibinfo {author} {\bibfnamefont {S.}~\bibnamefont {V{\'{e}}lez}},
  \bibinfo {author} {\bibfnamefont {J.~M.}\ \bibnamefont {Hern{\'{a}}ndez}},
  \bibinfo {author} {\bibfnamefont {D.}~\bibnamefont {Backes}}, \bibinfo
  {author} {\bibfnamefont {A.~D.}\ \bibnamefont {Kent}}, \ and\ \bibinfo
  {author} {\bibfnamefont {F.}~\bibnamefont {Maci{\`{a}}}},\ }\href {\doibase
  10.1103/PhysRevApplied.7.054027} {\bibfield  {journal} {\bibinfo  {journal}
  {Phys. Rev. Applied}\ }\textbf {\bibinfo {volume} {7}},\ \bibinfo {pages}
  {054027} (\bibinfo {year} {2017})}\BibitemShut {NoStop}%
\bibitem [{\citenamefont {Slobodianiuk}\ \emph {et~al.}(2017)\citenamefont
  {Slobodianiuk}, \citenamefont {Prokopenko},\ and\ \citenamefont
  {Melkov}}]{Slobodianiuk2017jmmm}%
  \BibitemOpen
  \bibfield  {author} {\bibinfo {author} {\bibfnamefont {D.}~\bibnamefont
  {Slobodianiuk}}, \bibinfo {author} {\bibfnamefont {O.}~\bibnamefont
  {Prokopenko}}, \ and\ \bibinfo {author} {\bibfnamefont {G.}~\bibnamefont
  {Melkov}},\ }\href {\doibase 10.1016/j.jmmm.2017.04.093} {\bibfield
  {journal} {\bibinfo  {journal} {J. Magn. Magn. Mater.}\ }\textbf {\bibinfo
  {volume} {439}},\ \bibinfo {pages} {144} (\bibinfo {year}
  {2017})}\BibitemShut {NoStop}%
\bibitem [{\citenamefont {Slavin}\ and\ \citenamefont
  {Tiberkevich}(2005)}]{Slavin2005}%
  \BibitemOpen
  \bibfield  {author} {\bibinfo {author} {\bibfnamefont {A.}~\bibnamefont
  {Slavin}}\ and\ \bibinfo {author} {\bibfnamefont {V.}~\bibnamefont
  {Tiberkevich}},\ }\href {\doibase 10.1103/PhysRevLett.95.237201} {\bibfield
  {journal} {\bibinfo  {journal} {Phys. Rev. Lett.}\ }\textbf {\bibinfo
  {volume} {95}},\ \bibinfo {pages} {237201} (\bibinfo {year}
  {2005})}\BibitemShut {NoStop}%
\bibitem [{\citenamefont {Bonetti}\ \emph {et~al.}(2010)\citenamefont
  {Bonetti}, \citenamefont {Tiberkevich}, \citenamefont {Consolo},
  \citenamefont {Finocchio}, \citenamefont {Muduli}, \citenamefont {Mancoff},
  \citenamefont {Slavin},\ and\ \citenamefont {\AA{}kerman}}]{Bonetti2010}%
  \BibitemOpen
  \bibfield  {author} {\bibinfo {author} {\bibfnamefont {S.}~\bibnamefont
  {Bonetti}}, \bibinfo {author} {\bibfnamefont {V.}~\bibnamefont
  {Tiberkevich}}, \bibinfo {author} {\bibfnamefont {G.}~\bibnamefont
  {Consolo}}, \bibinfo {author} {\bibfnamefont {G.}~\bibnamefont {Finocchio}},
  \bibinfo {author} {\bibfnamefont {P.}~\bibnamefont {Muduli}}, \bibinfo
  {author} {\bibfnamefont {F.}~\bibnamefont {Mancoff}}, \bibinfo {author}
  {\bibfnamefont {A.}~\bibnamefont {Slavin}}, \ and\ \bibinfo {author}
  {\bibfnamefont {J.}~\bibnamefont {\AA{}kerman}},\ }\href {\doibase
  10.1103/PhysRevLett.105.217204} {\bibfield  {journal} {\bibinfo  {journal}
  {Phys. Rev. Lett.}\ }\textbf {\bibinfo {volume} {105}},\ \bibinfo {pages}
  {217204} (\bibinfo {year} {2010})}\BibitemShut {NoStop}%
\bibitem [{\citenamefont {Demidov}\ \emph {et~al.}(2010)\citenamefont
  {Demidov}, \citenamefont {Urazhdin},\ and\ \citenamefont
  {Demokritov}}]{Demidov2010}%
  \BibitemOpen
  \bibfield  {author} {\bibinfo {author} {\bibfnamefont {V.~E.}\ \bibnamefont
  {Demidov}}, \bibinfo {author} {\bibfnamefont {S.}~\bibnamefont {Urazhdin}}, \
  and\ \bibinfo {author} {\bibfnamefont {S.~O.}\ \bibnamefont {Demokritov}},\
  }\href {\doibase 10.1038/nmat2882} {\bibfield  {journal} {\bibinfo  {journal}
  {Nat. Mater.}\ }\textbf {\bibinfo {volume} {9}},\ \bibinfo {pages} {984}
  (\bibinfo {year} {2010})}\BibitemShut {NoStop}%
\bibitem [{\citenamefont {Demidov}\ \emph {et~al.}(2012)\citenamefont
  {Demidov}, \citenamefont {Urazhdin}, \citenamefont {Ulrichs}, \citenamefont
  {Tiberkevich}, \citenamefont {Slavin}, \citenamefont {Baither}, \citenamefont
  {Schmitz},\ and\ \citenamefont {Demokritov}}]{Demidov2012}%
  \BibitemOpen
  \bibfield  {author} {\bibinfo {author} {\bibfnamefont {V.}~\bibnamefont
  {Demidov}}, \bibinfo {author} {\bibfnamefont {S.}~\bibnamefont {Urazhdin}},
  \bibinfo {author} {\bibfnamefont {H.}~\bibnamefont {Ulrichs}}, \bibinfo
  {author} {\bibfnamefont {V.}~\bibnamefont {Tiberkevich}}, \bibinfo {author}
  {\bibfnamefont {A.}~\bibnamefont {Slavin}}, \bibinfo {author} {\bibfnamefont
  {D.}~\bibnamefont {Baither}}, \bibinfo {author} {\bibfnamefont
  {G.}~\bibnamefont {Schmitz}}, \ and\ \bibinfo {author} {\bibfnamefont
  {S.}~\bibnamefont {Demokritov}},\ }\href {\doibase 10.1038/NMAT3459}
  {\bibfield  {journal} {\bibinfo  {journal} {Nat. Mater.}\ }\textbf {\bibinfo
  {volume} {11}},\ \bibinfo {pages} {1028} (\bibinfo {year}
  {2012})}\BibitemShut {NoStop}%
\bibitem [{\citenamefont {Bonetti}\ \emph {et~al.}(2012)\citenamefont
  {Bonetti}, \citenamefont {Puliafito}, \citenamefont {Consolo}, \citenamefont
  {Tiberkevich}, \citenamefont {Slavin},\ and\ \citenamefont
  {\AA{}kerman}}]{Bonetti2012}%
  \BibitemOpen
  \bibfield  {author} {\bibinfo {author} {\bibfnamefont {S.}~\bibnamefont
  {Bonetti}}, \bibinfo {author} {\bibfnamefont {V.}~\bibnamefont {Puliafito}},
  \bibinfo {author} {\bibfnamefont {G.}~\bibnamefont {Consolo}}, \bibinfo
  {author} {\bibfnamefont {V.~S.}\ \bibnamefont {Tiberkevich}}, \bibinfo
  {author} {\bibfnamefont {A.~N.}\ \bibnamefont {Slavin}}, \ and\ \bibinfo
  {author} {\bibfnamefont {J.}~\bibnamefont {\AA{}kerman}},\ }\href {\doibase
  10.1103/PhysRevB.85.174427} {\bibfield  {journal} {\bibinfo  {journal} {Phys.
  Rev. B}\ }\textbf {\bibinfo {volume} {85}},\ \bibinfo {pages} {174427}
  (\bibinfo {year} {2012})}\BibitemShut {NoStop}%
\bibitem [{\citenamefont {Dumas}\ \emph {et~al.}(2013)\citenamefont {Dumas},
  \citenamefont {Iacocca}, \citenamefont {Bonetti}, \citenamefont {Sani},
  \citenamefont {Mohseni}, \citenamefont {Eklund}, \citenamefont {Persson},
  \citenamefont {Heinonen},\ and\ \citenamefont {\AA{}kerman}}]{Dumas2013}%
  \BibitemOpen
  \bibfield  {author} {\bibinfo {author} {\bibfnamefont {R.~K.}\ \bibnamefont
  {Dumas}}, \bibinfo {author} {\bibfnamefont {E.}~\bibnamefont {Iacocca}},
  \bibinfo {author} {\bibfnamefont {S.}~\bibnamefont {Bonetti}}, \bibinfo
  {author} {\bibfnamefont {S.~R.}\ \bibnamefont {Sani}}, \bibinfo {author}
  {\bibfnamefont {S.~M.}\ \bibnamefont {Mohseni}}, \bibinfo {author}
  {\bibfnamefont {A.}~\bibnamefont {Eklund}}, \bibinfo {author} {\bibfnamefont
  {J.}~\bibnamefont {Persson}}, \bibinfo {author} {\bibfnamefont
  {O.}~\bibnamefont {Heinonen}}, \ and\ \bibinfo {author} {\bibfnamefont
  {J.}~\bibnamefont {\AA{}kerman}},\ }\href {\doibase
  10.1103/PhysRevLett.110.257202} {\bibfield  {journal} {\bibinfo  {journal}
  {Phys. Rev. Lett.}\ }\textbf {\bibinfo {volume} {110}},\ \bibinfo {pages}
  {257202} (\bibinfo {year} {2013})}\BibitemShut {NoStop}%
\bibitem [{\citenamefont {Slonczewski}(1996)}]{Slonczewski1996}%
  \BibitemOpen
  \bibfield  {author} {\bibinfo {author} {\bibfnamefont {J.~C.}\ \bibnamefont
  {Slonczewski}},\ }\href {\doibase DOI: 10.1016/0304-8853(96)00062-5}
  {\bibfield  {journal} {\bibinfo  {journal} {J. Magn. Magn. Mater.}\ }\textbf
  {\bibinfo {volume} {159}},\ \bibinfo {pages} {L1 } (\bibinfo {year}
  {1996})}\BibitemShut {NoStop}%
\bibitem [{\citenamefont {Berger}(1996)}]{Berger1996}%
  \BibitemOpen
  \bibfield  {author} {\bibinfo {author} {\bibfnamefont {L.}~\bibnamefont
  {Berger}},\ }\href {\doibase 10.1103/PhysRevB.54.9353} {\bibfield  {journal}
  {\bibinfo  {journal} {Phys. Rev. B}\ }\textbf {\bibinfo {volume} {54}},\
  \bibinfo {pages} {9353} (\bibinfo {year} {1996})}\BibitemShut {NoStop}%
\bibitem [{\citenamefont {Slonczewski}(1999)}]{Slonczewski1999}%
  \BibitemOpen
  \bibfield  {author} {\bibinfo {author} {\bibfnamefont {J.~C.}\ \bibnamefont
  {Slonczewski}},\ }\href {\doibase DOI: 10.1016/S0304-8853(99)00043-8}
  {\bibfield  {journal} {\bibinfo  {journal} {J. Magn. Magn. Mater.}\ }\textbf
  {\bibinfo {volume} {195}},\ \bibinfo {pages} {261 } (\bibinfo {year}
  {1999})}\BibitemShut {NoStop}%
\bibitem [{\citenamefont {Raabe}\ \emph {et~al.}(2000)\citenamefont {Raabe},
  \citenamefont {Pulwey}, \citenamefont {Sattler}, \citenamefont
  {Schweinböck}, \citenamefont {Zweck},\ and\ \citenamefont
  {Weiss}}]{Raabe2000}%
  \BibitemOpen
  \bibfield  {author} {\bibinfo {author} {\bibfnamefont {J.}~\bibnamefont
  {Raabe}}, \bibinfo {author} {\bibfnamefont {R.}~\bibnamefont {Pulwey}},
  \bibinfo {author} {\bibfnamefont {R.}~\bibnamefont {Sattler}}, \bibinfo
  {author} {\bibfnamefont {T.}~\bibnamefont {Schweinböck}}, \bibinfo {author}
  {\bibfnamefont {J.}~\bibnamefont {Zweck}}, \ and\ \bibinfo {author}
  {\bibfnamefont {D.}~\bibnamefont {Weiss}},\ }\href {\doibase
  10.1063/1.1289216} {\bibfield  {journal} {\bibinfo  {journal} {J. Appl.
  Phys.}\ }\textbf {\bibinfo {volume} {88}},\ \bibinfo {pages} {4437} (\bibinfo
  {year} {2000})}\BibitemShut {NoStop}%
\bibitem [{\citenamefont {Shinjo}\ \emph {et~al.}(2000)\citenamefont {Shinjo},
  \citenamefont {Okuno}, \citenamefont {Hassdorf}, \citenamefont {Shigeto},\
  and\ \citenamefont {Ono}}]{Shinjo2000}%
  \BibitemOpen
  \bibfield  {author} {\bibinfo {author} {\bibfnamefont {T.}~\bibnamefont
  {Shinjo}}, \bibinfo {author} {\bibfnamefont {T.}~\bibnamefont {Okuno}},
  \bibinfo {author} {\bibfnamefont {R.}~\bibnamefont {Hassdorf}}, \bibinfo
  {author} {\bibfnamefont {K.}~\bibnamefont {Shigeto}}, \ and\ \bibinfo
  {author} {\bibfnamefont {T.}~\bibnamefont {Ono}},\ }\href {\doibase
  10.1126/science.289.5481.930} {\bibfield  {journal} {\bibinfo  {journal}
  {Science}\ }\textbf {\bibinfo {volume} {289}},\ \bibinfo {pages} {930}
  (\bibinfo {year} {2000})}\BibitemShut {NoStop}%
\bibitem [{\citenamefont {R\"{o}{\ss}ler}\ \emph {et~al.}(2006)\citenamefont
  {R\"{o}{\ss}ler}, \citenamefont {Bogdanov},\ and\ \citenamefont
  {Pfleiderer}}]{Rossler2006}%
  \BibitemOpen
  \bibfield  {author} {\bibinfo {author} {\bibfnamefont {U.~K.}\ \bibnamefont
  {R\"{o}{\ss}ler}}, \bibinfo {author} {\bibfnamefont {A.~N.}\ \bibnamefont
  {Bogdanov}}, \ and\ \bibinfo {author} {\bibfnamefont {C.}~\bibnamefont
  {Pfleiderer}},\ }\href {\doibase 10.1038/nature05056} {\bibfield  {journal}
  {\bibinfo  {journal} {Nature}\ }\textbf {\bibinfo {volume} {442}},\ \bibinfo
  {pages} {797} (\bibinfo {year} {2006})}\BibitemShut {NoStop}%
\bibitem [{\citenamefont {Muhlbauer}\ \emph {et~al.}(2009)\citenamefont
  {Muhlbauer}, \citenamefont {Binz}, \citenamefont {Jonietz}, \citenamefont
  {Pfleiderer}, \citenamefont {Rosch}, \citenamefont {Neubauer}, \citenamefont
  {Georgii},\ and\ \citenamefont {Boni}}]{Pfleiderer2009}%
  \BibitemOpen
  \bibfield  {author} {\bibinfo {author} {\bibfnamefont {S.}~\bibnamefont
  {Muhlbauer}}, \bibinfo {author} {\bibfnamefont {B.}~\bibnamefont {Binz}},
  \bibinfo {author} {\bibfnamefont {F.}~\bibnamefont {Jonietz}}, \bibinfo
  {author} {\bibfnamefont {C.}~\bibnamefont {Pfleiderer}}, \bibinfo {author}
  {\bibfnamefont {A.}~\bibnamefont {Rosch}}, \bibinfo {author} {\bibfnamefont
  {A.}~\bibnamefont {Neubauer}}, \bibinfo {author} {\bibfnamefont
  {R.}~\bibnamefont {Georgii}}, \ and\ \bibinfo {author} {\bibfnamefont
  {P.}~\bibnamefont {Boni}},\ }\href {\doibase 10.1126/science.1166767}
  {\bibfield  {journal} {\bibinfo  {journal} {Science}\ }\textbf {\bibinfo
  {volume} {323}},\ \bibinfo {pages} {915} (\bibinfo {year}
  {2009})}\BibitemShut {NoStop}%
\bibitem [{\citenamefont {Yu}\ \emph {et~al.}(2010)\citenamefont {Yu},
  \citenamefont {Onose}, \citenamefont {Kanazawa}, \citenamefont {Park},
  \citenamefont {Han}, \citenamefont {Matsui}, \citenamefont {Nagaosa},\ and\
  \citenamefont {Tokura}}]{Yu2010b}%
  \BibitemOpen
  \bibfield  {author} {\bibinfo {author} {\bibfnamefont {X.~Z.}\ \bibnamefont
  {Yu}}, \bibinfo {author} {\bibfnamefont {Y.}~\bibnamefont {Onose}}, \bibinfo
  {author} {\bibfnamefont {N.}~\bibnamefont {Kanazawa}}, \bibinfo {author}
  {\bibfnamefont {J.~H.}\ \bibnamefont {Park}}, \bibinfo {author}
  {\bibfnamefont {J.~H.}\ \bibnamefont {Han}}, \bibinfo {author} {\bibfnamefont
  {Y.}~\bibnamefont {Matsui}}, \bibinfo {author} {\bibfnamefont
  {N.}~\bibnamefont {Nagaosa}}, \ and\ \bibinfo {author} {\bibfnamefont
  {Y.}~\bibnamefont {Tokura}},\ }\href {\doibase 10.1038/nature09124}
  {\bibfield  {journal} {\bibinfo  {journal} {Nature}\ }\textbf {\bibinfo
  {volume} {465}},\ \bibinfo {pages} {901} (\bibinfo {year}
  {2010})}\BibitemShut {NoStop}%
\bibitem [{\citenamefont {Romming}\ \emph {et~al.}(2013)\citenamefont
  {Romming}, \citenamefont {Hanneken}, \citenamefont {Menzel}, \citenamefont
  {Bickel}, \citenamefont {Wolter}, \citenamefont {von Bergmann}, \citenamefont
  {Kubetzka},\ and\ \citenamefont {Wiesendanger}}]{Romming2013}%
  \BibitemOpen
  \bibfield  {author} {\bibinfo {author} {\bibfnamefont {N.}~\bibnamefont
  {Romming}}, \bibinfo {author} {\bibfnamefont {C.}~\bibnamefont {Hanneken}},
  \bibinfo {author} {\bibfnamefont {M.}~\bibnamefont {Menzel}}, \bibinfo
  {author} {\bibfnamefont {J.~E.}\ \bibnamefont {Bickel}}, \bibinfo {author}
  {\bibfnamefont {B.}~\bibnamefont {Wolter}}, \bibinfo {author} {\bibfnamefont
  {K.}~\bibnamefont {von Bergmann}}, \bibinfo {author} {\bibfnamefont
  {a.}~\bibnamefont {Kubetzka}}, \ and\ \bibinfo {author} {\bibfnamefont
  {R.}~\bibnamefont {Wiesendanger}},\ }\href {\doibase 10.1126/science.1240573}
  {\bibfield  {journal} {\bibinfo  {journal} {Science}\ }\textbf {\bibinfo
  {volume} {341}},\ \bibinfo {pages} {636} (\bibinfo {year}
  {2013})}\BibitemShut {NoStop}%
\bibitem [{\citenamefont {Zhou}\ \emph {et~al.}(2015)\citenamefont {Zhou},
  \citenamefont {Iacocca}, \citenamefont {Awad}, \citenamefont {Dumas},
  \citenamefont {Zhang}, \citenamefont {Braun},\ and\ \citenamefont
  {\AA~kerman}}]{Zhou2015}%
  \BibitemOpen
  \bibfield  {author} {\bibinfo {author} {\bibfnamefont {Y.}~\bibnamefont
  {Zhou}}, \bibinfo {author} {\bibfnamefont {E.}~\bibnamefont {Iacocca}},
  \bibinfo {author} {\bibfnamefont {A.~A.}\ \bibnamefont {Awad}}, \bibinfo
  {author} {\bibfnamefont {R.~K.}\ \bibnamefont {Dumas}}, \bibinfo {author}
  {\bibfnamefont {F.~C.}\ \bibnamefont {Zhang}}, \bibinfo {author}
  {\bibfnamefont {H.~B.}\ \bibnamefont {Braun}}, \ and\ \bibinfo {author}
  {\bibfnamefont {J.}~\bibnamefont {\AA~kerman}},\ }\href {\doibase
  10.1038/ncomms9193} {\bibfield  {journal} {\bibinfo  {journal} {Nat.
  Commun.}\ }\textbf {\bibinfo {volume} {6}},\ \bibinfo {pages} {8193}
  (\bibinfo {year} {2015})}\BibitemShut {NoStop}%
\bibitem [{\citenamefont {Liu}\ \emph {et~al.}(2015)\citenamefont {Liu},
  \citenamefont {Lim},\ and\ \citenamefont {Urazhdin}}]{Liu2015b}%
  \BibitemOpen
  \bibfield  {author} {\bibinfo {author} {\bibfnamefont {R.~H.}\ \bibnamefont
  {Liu}}, \bibinfo {author} {\bibfnamefont {W.~L.}\ \bibnamefont {Lim}}, \ and\
  \bibinfo {author} {\bibfnamefont {S.}~\bibnamefont {Urazhdin}},\ }\href
  {\doibase 10.1103/PhysRevLett.114.137201} {\bibfield  {journal} {\bibinfo
  {journal} {Phys. Rev. Lett.}\ }\textbf {\bibinfo {volume} {114}},\ \bibinfo
  {pages} {137201} (\bibinfo {year} {2015})}\BibitemShut {NoStop}%
\bibitem [{\citenamefont {Dumas}\ \emph {et~al.}(2014)\citenamefont {Dumas},
  \citenamefont {Sani}, \citenamefont {Mohseni}, \citenamefont {Iacocca},
  \citenamefont {Pogoryelov}, \citenamefont {Muduli}, \citenamefont {Chung},
  \citenamefont {Dürrenfeld},\ and\ \citenamefont {Åkerman}}]{Dumas2014}%
  \BibitemOpen
  \bibfield  {author} {\bibinfo {author} {\bibfnamefont {R.}~\bibnamefont
  {Dumas}}, \bibinfo {author} {\bibfnamefont {S.}~\bibnamefont {Sani}},
  \bibinfo {author} {\bibfnamefont {S.}~\bibnamefont {Mohseni}}, \bibinfo
  {author} {\bibfnamefont {E.}~\bibnamefont {Iacocca}}, \bibinfo {author}
  {\bibfnamefont {Y.}~\bibnamefont {Pogoryelov}}, \bibinfo {author}
  {\bibfnamefont {P.}~\bibnamefont {Muduli}}, \bibinfo {author} {\bibfnamefont
  {S.}~\bibnamefont {Chung}}, \bibinfo {author} {\bibfnamefont
  {P.}~\bibnamefont {Dürrenfeld}}, \ and\ \bibinfo {author} {\bibfnamefont
  {J.}~\bibnamefont {Åkerman}},\ }\href {\doibase 10.1109/TMAG.2014.2305762}
  {\bibfield  {journal} {\bibinfo  {journal} {IEEE Trans. Magn.}\ }\textbf
  {\bibinfo {volume} {50}},\ \bibinfo {pages} {257202} (\bibinfo {year}
  {2014})}\BibitemShut {NoStop}%
\bibitem [{\citenamefont {Chen}\ \emph {et~al.}(2016)\citenamefont {Chen},
  \citenamefont {Dumas}, \citenamefont {Eklund}, \citenamefont {Muduli},
  \citenamefont {Houshang}, \citenamefont {Awad}, \citenamefont {Durrenfeld},
  \citenamefont {Malm}, \citenamefont {Rusu},\ and\ \citenamefont
  {Akerman}}]{Chen2016procieee}%
  \BibitemOpen
  \bibfield  {author} {\bibinfo {author} {\bibfnamefont {T.}~\bibnamefont
  {Chen}}, \bibinfo {author} {\bibfnamefont {R.~K.}\ \bibnamefont {Dumas}},
  \bibinfo {author} {\bibfnamefont {A.}~\bibnamefont {Eklund}}, \bibinfo
  {author} {\bibfnamefont {P.~K.}\ \bibnamefont {Muduli}}, \bibinfo {author}
  {\bibfnamefont {A.}~\bibnamefont {Houshang}}, \bibinfo {author}
  {\bibfnamefont {A.~A.}\ \bibnamefont {Awad}}, \bibinfo {author}
  {\bibfnamefont {P.}~\bibnamefont {Durrenfeld}}, \bibinfo {author}
  {\bibfnamefont {B.~G.}\ \bibnamefont {Malm}}, \bibinfo {author}
  {\bibfnamefont {A.}~\bibnamefont {Rusu}}, \ and\ \bibinfo {author}
  {\bibfnamefont {J.}~\bibnamefont {Akerman}},\ }\href {\doibase
  10.1109/JPROC.2016.2554518} {\bibfield  {journal} {\bibinfo  {journal}
  {Proceedings of the IEEE}\ }\textbf {\bibinfo {volume} {104}},\ \bibinfo
  {pages} {1919} (\bibinfo {year} {2016})}\BibitemShut {NoStop}%
\bibitem [{\citenamefont {{T}soi}\ \emph {et~al.}(1998)\citenamefont {{T}soi},
  \citenamefont {{J}ansen}, \citenamefont {{B}ass}, \citenamefont {{C}hiang},
  \citenamefont {{S}eck}, \citenamefont {{T}soi},\ and\ \citenamefont
  {{W}yder}}]{tsoi1998prl}%
  \BibitemOpen
  \bibfield  {author} {\bibinfo {author} {\bibfnamefont {M.}~\bibnamefont
  {{T}soi}}, \bibinfo {author} {\bibfnamefont {A.~G.~M.}\ \bibnamefont
  {{J}ansen}}, \bibinfo {author} {\bibfnamefont {J.}~\bibnamefont {{B}ass}},
  \bibinfo {author} {\bibfnamefont {W.-C.}\ \bibnamefont {{C}hiang}}, \bibinfo
  {author} {\bibfnamefont {M.}~\bibnamefont {{S}eck}}, \bibinfo {author}
  {\bibfnamefont {V.}~\bibnamefont {{T}soi}}, \ and\ \bibinfo {author}
  {\bibfnamefont {P.}~\bibnamefont {{W}yder}},\ }\href {\doibase
  10.1103/PhysRevLett.80.4281} {\bibfield  {journal} {\bibinfo  {journal}
  {Phys. Rev. Lett.}\ }\textbf {\bibinfo {volume} {80}},\ \bibinfo {pages}
  {4281} (\bibinfo {year} {1998})}\BibitemShut {NoStop}%
\bibitem [{\citenamefont {{Tsoi}}\ \emph {et~al.}(2000)\citenamefont {{Tsoi}},
  \citenamefont {{Jansen}}, \citenamefont {{Bass}}, \citenamefont {{Chiang}},
  \citenamefont {{Tsoi}},\ and\ \citenamefont {{Wyder}}}]{tsoi2000nt}%
  \BibitemOpen
  \bibfield  {author} {\bibinfo {author} {\bibfnamefont {M.}~\bibnamefont
  {{Tsoi}}}, \bibinfo {author} {\bibfnamefont {A.~G.~M.}\ \bibnamefont
  {{Jansen}}}, \bibinfo {author} {\bibfnamefont {J.}~\bibnamefont {{Bass}}},
  \bibinfo {author} {\bibfnamefont {W.-C.}\ \bibnamefont {{Chiang}}}, \bibinfo
  {author} {\bibfnamefont {V.}~\bibnamefont {{Tsoi}}}, \ and\ \bibinfo {author}
  {\bibfnamefont {P.}~\bibnamefont {{Wyder}}},\ }\href {\doibase
  10.1038/35017512} {\bibfield  {journal} {\bibinfo  {journal} {Nature}\
  }\textbf {\bibinfo {volume} {406}},\ \bibinfo {pages} {46} (\bibinfo {year}
  {2000})}\BibitemShut {NoStop}%
\bibitem [{\citenamefont {{R}ippard}\ \emph {et~al.}(2004)\citenamefont
  {{R}ippard}, \citenamefont {{P}ufall}, \citenamefont {{K}aka}, \citenamefont
  {{R}ussek},\ and\ \citenamefont {{S}ilva}}]{rippard2004prl}%
  \BibitemOpen
  \bibfield  {author} {\bibinfo {author} {\bibfnamefont {W.}~\bibnamefont
  {{R}ippard}}, \bibinfo {author} {\bibfnamefont {M.}~\bibnamefont {{P}ufall}},
  \bibinfo {author} {\bibfnamefont {S.}~\bibnamefont {{K}aka}}, \bibinfo
  {author} {\bibfnamefont {S.}~\bibnamefont {{R}ussek}}, \ and\ \bibinfo
  {author} {\bibfnamefont {T.}~\bibnamefont {{S}ilva}},\ }\href {\doibase
  10.1103/PhysRevLett.92.027201} {\bibfield  {journal} {\bibinfo  {journal}
  {Phys. Rev. Lett.}\ }\textbf {\bibinfo {volume} {92}},\ \bibinfo {pages}
  {027201} (\bibinfo {year} {2004})}\BibitemShut {NoStop}%
\bibitem [{\citenamefont {Slavin}\ and\ \citenamefont
  {Kabos}(2005)}]{SlavinKabos2005}%
  \BibitemOpen
  \bibfield  {author} {\bibinfo {author} {\bibfnamefont {A.}~\bibnamefont
  {Slavin}}\ and\ \bibinfo {author} {\bibfnamefont {P.}~\bibnamefont {Kabos}},\
  }\href {\doibase 10.1109/TMAG.2005.845915} {\bibfield  {journal} {\bibinfo
  {journal} {IEEE Trans. Magn.}\ }\textbf {\bibinfo {volume} {41}},\ \bibinfo
  {pages} {1264} (\bibinfo {year} {2005})}\BibitemShut {NoStop}%
\bibitem [{\citenamefont {Slavin}\ and\ \citenamefont
  {Tiberkevich}(2009)}]{Slavin2009}%
  \BibitemOpen
  \bibfield  {author} {\bibinfo {author} {\bibfnamefont {A.}~\bibnamefont
  {Slavin}}\ and\ \bibinfo {author} {\bibfnamefont {V.}~\bibnamefont
  {Tiberkevich}},\ }\href {\doibase 10.1109/TMAG.2008.2009935} {\bibfield
  {journal} {\bibinfo  {journal} {IEEE Trans. Magn.}\ }\textbf {\bibinfo
  {volume} {45}},\ \bibinfo {pages} {1875 } (\bibinfo {year}
  {2009})}\BibitemShut {NoStop}%
\bibitem [{\citenamefont {Hoefer}\ \emph {et~al.}(2008)\citenamefont {Hoefer},
  \citenamefont {Silva},\ and\ \citenamefont {Stiles}}]{Hoefer2008prb}%
  \BibitemOpen
  \bibfield  {author} {\bibinfo {author} {\bibfnamefont {M.}~\bibnamefont
  {Hoefer}}, \bibinfo {author} {\bibfnamefont {T.}~\bibnamefont {Silva}}, \
  and\ \bibinfo {author} {\bibfnamefont {M.}~\bibnamefont {Stiles}},\ }\href
  {http://prb.aps.org/abstract/PRB/v77/i14/e144401} {\bibfield  {journal}
  {\bibinfo  {journal} {Phys. Rev. B}\ }\textbf {\bibinfo {volume} {77}},\
  \bibinfo {pages} {144401} (\bibinfo {year} {2008})}\BibitemShut {NoStop}%
\bibitem [{\citenamefont {Madami}\ \emph {et~al.}(2015)\citenamefont {Madami},
  \citenamefont {Iacocca}, \citenamefont {Sani}, \citenamefont {Gubbiotti},
  \citenamefont {Tacchi}, \citenamefont {Dumas}, \citenamefont {{\AA}kerman},\
  and\ \citenamefont {Carlotti}}]{Madami2015prb}%
  \BibitemOpen
  \bibfield  {author} {\bibinfo {author} {\bibfnamefont {M.}~\bibnamefont
  {Madami}}, \bibinfo {author} {\bibfnamefont {E.}~\bibnamefont {Iacocca}},
  \bibinfo {author} {\bibfnamefont {S.}~\bibnamefont {Sani}}, \bibinfo {author}
  {\bibfnamefont {G.}~\bibnamefont {Gubbiotti}}, \bibinfo {author}
  {\bibfnamefont {S.}~\bibnamefont {Tacchi}}, \bibinfo {author} {\bibfnamefont
  {R.~K.}\ \bibnamefont {Dumas}}, \bibinfo {author} {\bibfnamefont
  {J.}~\bibnamefont {{\AA}kerman}}, \ and\ \bibinfo {author} {\bibfnamefont
  {G.}~\bibnamefont {Carlotti}},\ }\href {\doibase 10.1103/PhysRevB.92.024403}
  {\bibfield  {journal} {\bibinfo  {journal} {Phys. Rev. B}\ }\textbf {\bibinfo
  {volume} {92}},\ \bibinfo {pages} {024403} (\bibinfo {year}
  {2015})}\BibitemShut {NoStop}%
\bibitem [{\citenamefont {Houshang}\ \emph {et~al.}(2016)\citenamefont
  {Houshang}, \citenamefont {Iacocca}, \citenamefont {D{\"{u}}rrenfeld},
  \citenamefont {Sani}, \citenamefont {{\AA}kerman},\ and\ \citenamefont
  {Dumas}}]{Houshang2016natnano}%
  \BibitemOpen
  \bibfield  {author} {\bibinfo {author} {\bibfnamefont {A.}~\bibnamefont
  {Houshang}}, \bibinfo {author} {\bibfnamefont {E.}~\bibnamefont {Iacocca}},
  \bibinfo {author} {\bibfnamefont {P.}~\bibnamefont {D{\"{u}}rrenfeld}},
  \bibinfo {author} {\bibfnamefont {S.~R.}\ \bibnamefont {Sani}}, \bibinfo
  {author} {\bibfnamefont {J.}~\bibnamefont {{\AA}kerman}}, \ and\ \bibinfo
  {author} {\bibfnamefont {R.~K.}\ \bibnamefont {Dumas}},\ }\href {\doibase
  10.1038/nnano.2015.280} {\bibfield  {journal} {\bibinfo  {journal} {Nat.
  Nanotechnol.}\ }\textbf {\bibinfo {volume} {11}},\ \bibinfo {pages} {280}
  (\bibinfo {year} {2016})}\BibitemShut {NoStop}%
\bibitem [{\citenamefont {{Kaka}}\ \emph {et~al.}(2005)\citenamefont {{Kaka}},
  \citenamefont {{Pufall}}, \citenamefont {{Rippard}}, \citenamefont {{Silva}},
  \citenamefont {{Russek}},\ and\ \citenamefont {{Katine}}}]{kaka2005nt}%
  \BibitemOpen
  \bibfield  {author} {\bibinfo {author} {\bibfnamefont {S.}~\bibnamefont
  {{Kaka}}}, \bibinfo {author} {\bibfnamefont {M.~R.}\ \bibnamefont
  {{Pufall}}}, \bibinfo {author} {\bibfnamefont {W.~H.}\ \bibnamefont
  {{Rippard}}}, \bibinfo {author} {\bibfnamefont {T.~J.}\ \bibnamefont
  {{Silva}}}, \bibinfo {author} {\bibfnamefont {S.~E.}\ \bibnamefont
  {{Russek}}}, \ and\ \bibinfo {author} {\bibfnamefont {J.~A.}\ \bibnamefont
  {{Katine}}},\ }\href {\doibase 10.1038/nature04035} {\bibfield  {journal}
  {\bibinfo  {journal} {Nature}\ }\textbf {\bibinfo {volume} {437}},\ \bibinfo
  {pages} {389} (\bibinfo {year} {2005})}\BibitemShut {NoStop}%
\bibitem [{\citenamefont {{Mancoff}}\ \emph {et~al.}(2005)\citenamefont
  {{Mancoff}}, \citenamefont {{Rizzo}}, \citenamefont {{Engel}},\ and\
  \citenamefont {{Tehrani}}}]{mancoff2005nt}%
  \BibitemOpen
  \bibfield  {author} {\bibinfo {author} {\bibfnamefont {F.~B.}\ \bibnamefont
  {{Mancoff}}}, \bibinfo {author} {\bibfnamefont {N.~D.}\ \bibnamefont
  {{Rizzo}}}, \bibinfo {author} {\bibfnamefont {B.~N.}\ \bibnamefont
  {{Engel}}}, \ and\ \bibinfo {author} {\bibfnamefont {S.}~\bibnamefont
  {{Tehrani}}},\ }\href {\doibase 10.1038/nature04036} {\bibfield  {journal}
  {\bibinfo  {journal} {Nature}\ }\textbf {\bibinfo {volume} {437}},\ \bibinfo
  {pages} {393} (\bibinfo {year} {2005})}\BibitemShut {NoStop}%
\bibitem [{\citenamefont {Sani}\ \emph {et~al.}(2013)\citenamefont {Sani},
  \citenamefont {Persson}, \citenamefont {Mohseni}, \citenamefont {Pogoryelov},
  \citenamefont {Muduli}, \citenamefont {Eklund}, \citenamefont {Malm},
  \citenamefont {K{\"{a}}ll}, \citenamefont {Dmitriev},\ and\ \citenamefont
  {{\AA}kerman}}]{sani2013ntc}%
  \BibitemOpen
  \bibfield  {author} {\bibinfo {author} {\bibfnamefont {S.}~\bibnamefont
  {Sani}}, \bibinfo {author} {\bibfnamefont {J.}~\bibnamefont {Persson}},
  \bibinfo {author} {\bibfnamefont {S.~M.}\ \bibnamefont {Mohseni}}, \bibinfo
  {author} {\bibfnamefont {Y.}~\bibnamefont {Pogoryelov}}, \bibinfo {author}
  {\bibfnamefont {P.~K.}\ \bibnamefont {Muduli}}, \bibinfo {author}
  {\bibfnamefont {A.}~\bibnamefont {Eklund}}, \bibinfo {author} {\bibfnamefont
  {G.}~\bibnamefont {Malm}}, \bibinfo {author} {\bibfnamefont {M.}~\bibnamefont
  {K{\"{a}}ll}}, \bibinfo {author} {\bibfnamefont {A.}~\bibnamefont
  {Dmitriev}}, \ and\ \bibinfo {author} {\bibfnamefont {J.}~\bibnamefont
  {{\AA}kerman}},\ }\href {\doibase 10.1038/ncomms3731} {\bibfield  {journal}
  {\bibinfo  {journal} {Nat. Commun.}\ }\textbf {\bibinfo {volume} {4}},\
  \bibinfo {pages} {2731} (\bibinfo {year} {2013})}\BibitemShut {NoStop}%
\bibitem [{\citenamefont {Rippard}\ \emph {et~al.}(2010)\citenamefont
  {Rippard}, \citenamefont {Deac}, \citenamefont {Pufall}, \citenamefont
  {Shaw}, \citenamefont {Keller}, \citenamefont {Russek},\ and\ \citenamefont
  {Serpico}}]{Rippard2010prb}%
  \BibitemOpen
  \bibfield  {author} {\bibinfo {author} {\bibfnamefont {W.~H.}\ \bibnamefont
  {Rippard}}, \bibinfo {author} {\bibfnamefont {A.~M.}\ \bibnamefont {Deac}},
  \bibinfo {author} {\bibfnamefont {M.~R.}\ \bibnamefont {Pufall}}, \bibinfo
  {author} {\bibfnamefont {J.~M.}\ \bibnamefont {Shaw}}, \bibinfo {author}
  {\bibfnamefont {M.~W.}\ \bibnamefont {Keller}}, \bibinfo {author}
  {\bibfnamefont {S.~E.}\ \bibnamefont {Russek}}, \ and\ \bibinfo {author}
  {\bibfnamefont {C.}~\bibnamefont {Serpico}},\ }\href {\doibase
  10.1103/PhysRevB.81.014426} {\bibfield  {journal} {\bibinfo  {journal} {Phys.
  Rev. B}\ }\textbf {\bibinfo {volume} {81}},\ \bibinfo {pages} {014426}
  (\bibinfo {year} {2010})}\BibitemShut {NoStop}%
\bibitem [{\citenamefont {Mohseni}\ \emph {et~al.}(2011)\citenamefont
  {Mohseni}, \citenamefont {Sani}, \citenamefont {Persson}, \citenamefont
  {Anh~Nguyen}, \citenamefont {Chung}, \citenamefont {Pogoryelov},\ and\
  \citenamefont {\AA{}kerman}}]{Mohseni2011}%
  \BibitemOpen
  \bibfield  {author} {\bibinfo {author} {\bibfnamefont {S.~M.}\ \bibnamefont
  {Mohseni}}, \bibinfo {author} {\bibfnamefont {S.~R.}\ \bibnamefont {Sani}},
  \bibinfo {author} {\bibfnamefont {J.}~\bibnamefont {Persson}}, \bibinfo
  {author} {\bibfnamefont {T.~N.}\ \bibnamefont {Anh~Nguyen}}, \bibinfo
  {author} {\bibfnamefont {S.}~\bibnamefont {Chung}}, \bibinfo {author}
  {\bibfnamefont {Y.}~\bibnamefont {Pogoryelov}}, \ and\ \bibinfo {author}
  {\bibfnamefont {J.}~\bibnamefont {\AA{}kerman}},\ }\href {\doibase
  10.1002/pssr.201105375} {\bibfield  {journal} {\bibinfo  {journal} {Phys.
  Status Solidi RRL}\ }\textbf {\bibinfo {volume} {5}},\ \bibinfo {pages} {432
  } (\bibinfo {year} {2011})}\BibitemShut {NoStop}%
\bibitem [{\citenamefont {Backes}\ \emph {et~al.}(2015)\citenamefont {Backes},
  \citenamefont {Maci{\`{a}}}, \citenamefont {Bonetti}, \citenamefont
  {Kukreja}, \citenamefont {Ohldag},\ and\ \citenamefont {Kent}}]{Backes2015}%
  \BibitemOpen
  \bibfield  {author} {\bibinfo {author} {\bibfnamefont {D.}~\bibnamefont
  {Backes}}, \bibinfo {author} {\bibfnamefont {F.}~\bibnamefont {Maci{\`{a}}}},
  \bibinfo {author} {\bibfnamefont {S.}~\bibnamefont {Bonetti}}, \bibinfo
  {author} {\bibfnamefont {R.}~\bibnamefont {Kukreja}}, \bibinfo {author}
  {\bibfnamefont {H.}~\bibnamefont {Ohldag}}, \ and\ \bibinfo {author}
  {\bibfnamefont {A.~D.}\ \bibnamefont {Kent}},\ }\href {\doibase
  10.1103/PhysRevLett.115.127205} {\bibfield  {journal} {\bibinfo  {journal}
  {Phys. Rev. Lett.}\ }\textbf {\bibinfo {volume} {115}},\ \bibinfo {pages}
  {127205} (\bibinfo {year} {2015})}\BibitemShut {NoStop}%
\bibitem [{\citenamefont {Banuazizi}\ \emph {et~al.}(2017)\citenamefont
  {Banuazizi}, \citenamefont {Sani}, \citenamefont {Eklund}, \citenamefont
  {Naiini}, \citenamefont {Mohseni}, \citenamefont {Chung}, \citenamefont
  {D{\"{u}}rrenfeld}, \citenamefont {Malm},\ and\ \citenamefont
  {{\AA}kerman}}]{Banuazizi2017nanoscale}%
  \BibitemOpen
  \bibfield  {author} {\bibinfo {author} {\bibfnamefont {S.~A.~H.}\
  \bibnamefont {Banuazizi}}, \bibinfo {author} {\bibfnamefont {S.~R.}\
  \bibnamefont {Sani}}, \bibinfo {author} {\bibfnamefont {A.}~\bibnamefont
  {Eklund}}, \bibinfo {author} {\bibfnamefont {M.~M.}\ \bibnamefont {Naiini}},
  \bibinfo {author} {\bibfnamefont {S.~M.}\ \bibnamefont {Mohseni}}, \bibinfo
  {author} {\bibfnamefont {S.}~\bibnamefont {Chung}}, \bibinfo {author}
  {\bibfnamefont {P.}~\bibnamefont {D{\"{u}}rrenfeld}}, \bibinfo {author}
  {\bibfnamefont {B.~G.}\ \bibnamefont {Malm}}, \ and\ \bibinfo {author}
  {\bibfnamefont {J.}~\bibnamefont {{\AA}kerman}},\ }\href {\doibase
  10.1039/C6NR07309C} {\bibfield  {journal} {\bibinfo  {journal} {Nanoscale}\
  }\textbf {\bibinfo {volume} {9}},\ \bibinfo {pages} {1896} (\bibinfo {year}
  {2017})}\BibitemShut {NoStop}%
\bibitem [{\citenamefont {Nolle}\ \emph {et~al.}(2012)\citenamefont {Nolle},
  \citenamefont {Weigand}, \citenamefont {Audehm}, \citenamefont {Goering},
  \citenamefont {Wiesemann}, \citenamefont {Wolter}, \citenamefont {Nolle},\
  and\ \citenamefont {Schütz}}]{Nolle2012}%
  \BibitemOpen
  \bibfield  {author} {\bibinfo {author} {\bibfnamefont {D.}~\bibnamefont
  {Nolle}}, \bibinfo {author} {\bibfnamefont {M.}~\bibnamefont {Weigand}},
  \bibinfo {author} {\bibfnamefont {P.}~\bibnamefont {Audehm}}, \bibinfo
  {author} {\bibfnamefont {E.}~\bibnamefont {Goering}}, \bibinfo {author}
  {\bibfnamefont {U.}~\bibnamefont {Wiesemann}}, \bibinfo {author}
  {\bibfnamefont {C.}~\bibnamefont {Wolter}}, \bibinfo {author} {\bibfnamefont
  {E.}~\bibnamefont {Nolle}}, \ and\ \bibinfo {author} {\bibfnamefont
  {G.}~\bibnamefont {Schütz}},\ }\href {\doibase 10.1063/1.4707747} {\bibfield
   {journal} {\bibinfo  {journal} {Rev. Sci. Instrum.}\ }\textbf {\bibinfo
  {volume} {83}},\ \bibinfo {pages} {046112} (\bibinfo {year}
  {2012})}\BibitemShut {NoStop}%
\bibitem [{\citenamefont {Rasband}(2015)}]{ImageJ}%
  \BibitemOpen
  \bibfield  {author} {\bibinfo {author} {\bibfnamefont {W.}~\bibnamefont
  {Rasband}},\ }\href@noop {} {\enquote {\bibinfo {title} {{ImageJ}},}\ }
  (\bibinfo {year} {1997-2015})\BibitemShut {NoStop}%
\end{thebibliography}%

\end{document}